\documentclass[CRPHYS,Unicode,manuscript]{cedram}

\newcommand{\CORR}[1]{\textcolor{black}{#1}}

\title{Vibrations and Heat Transfer in Glasses: the role played by Disorder}
\alttitle{Vibrations et Transferts de chaleur dans les Verres: le rôle du Désordre}

\author{\firstname{Anne} \lastname{Tanguy}}
\address{Univ Lyon, INSA Lyon, CNRS, LaMCoS, UMR5259, Villeurbanne, 69621, France}
\address{ONERA, University Paris-Saclay, Chemin de la Hunière, BP 80100, Palaiseau, 92123, France}
\email[A. Tanguy]{tanguy.anne1@gmail.com}
\thanks{The author is funded by ANR fellowships DENSE - ANR-21-CE08-0005 and RATES - ANR-20-CE08-0022.  } 
\altkeywords{Matériaux amorphes, Verres, Atténuation acoustique, Propriétés thermiques, Comportement Thermo-Mécanique, Plasticité}
\begin{abstract} 
Amorphous materials are also distinguished from crystals by their thermal properties. The structural disorder seems to be responsible both for a significant increase in heat capacity compared to crystals of the same composition, but also for a significant decrease in thermal conductivity. The temperature dependence of thermal conductivity, unusual for common interpretations of solid-state physics, gave rise to a lot of debates. We review in this article different interpretations of thermal conductivity in amorphous materials. We show finally that the temperature dependence of thermal conductivity in dielectric materials can be understood by relating it to the disorder-dependent harmonic vibrational eigenmodes.
\end{abstract}
\begin{altabstract}
Les matériaux amorphes se distinguent aussi des cristaux par leurs propriétés thermiques. Le désordre structural semble être responsable à la fois d'une augmentation importante de la capacité calorifique par rapport aux cristaux de même composition, mais aussi d'une diminution importante de la conductivité thermique. La dépendance en température de la conductivité thermique, inhabituelle pour les interprétations usuelles de la physique du solide, a fait couler beaucoup d'encre. Nous passons en revue dans cet article différentes interprétations de la conductivité thermique dans les matériaux amorphes. Nous montrons que la dépendance en température dans les matériaux diéelectriques peut aussi être comprise à l'aide de l'effet du désordre structural sur les modes propres de vibration.
\end{altabstract}
\begin{document}

% Use the \maketitle command after the abstract
\maketitle

%% 2. French one if the paper is in English
% \selectlanguage{french}
% \section*{Version française abrégée}

%***

\selectlanguage{english}
%to go back to main language.

% Example of section
\section{Introduction}

Thermal properties of glasses are important to take into account, when they are used as building materials for energy saving and comfort for example. But the use of glasses as dielectric components in electrical circuits, makes heat management crucial as well to prevent the melting of the circuit submitted to high power fluxes ($\propto kW/cm^2$)~\cite{Oltmanns2020}. In the case of dielectric materials like silicate glasses, and also for semi-conductors like amorphous silicon, the heat is mainly supported by the atomic vibrations~\cite{Kittel2004,Tien1994}. This is why the current paper is focused on these two aspects: the thermal and the vibrational properties of glasses, or more generally of amorphous materials.

Amorphous materials are characterized by a fully disordered structure, with a lack of clear structural lengthscale, apart from the mean interatomic distance acting as a lower scale cutoff. In general, no structuration is visible at long lengthscales~\cite{Haberl2009, Tanguy2015}, but nanometric lengthscales have a signature, that can be measured mainly through the high frequency vibrational properties~\cite{Tanguy2002, Leonforte2006, Tanguy2015, Beltukov2016}. The existence and the possible role of medium-range order is also regularly discussed in relation to the ring structure in silica glasses~\cite{Rino1993} or to highly symetrical clusters~\cite{Tanaka2003} eventually with a fractal shape~\cite{Ma2008} or even to nanostrings~\cite{Bianchi2020, Tanaka2022}. Medium Range order has recently been identified in the dynamical structure factor~\cite{Egami2019}, thanks to its temperature sensitivity when approaching the glass transition temperature. When submitted to mechanical loads, glasses undergo plastic (dissipative) rearrangements at a sub-nanometric lengthscale~\cite{Maloney2004,Tanguy2006} corresponding to the core size of Eshelby inclusions~\cite{Albaret2016} also called shear transformation zones~\cite{Argon1979, Falk1998}. This last name is not the best, since these transformation zones combine indeed shear with hydrostatic compression~\cite{Albaret2016}. The plastic deformation is an anharmonic irreversible response refering as well to the free-volume theory~\cite{Cohen1959,Spaepen1979} based on the nucleation and diffusion of localized defects. However, a structural signature of such defects is very hard to find, in many glasses~\cite{Fusco2010, Patinet2020}. Only in silicate glasses in the plastic flow regime, plastic rearrangements in a sufficiently large number with a permanent structural signature has been evidenced thanks to their vibrational properties, using Raman spectroscopy~\cite{Martinet2020}. It has been interpreted with the help of Molecular Dynamics simulations as an irreversible modification of the ring structure giving rise to enhanced streching vibrations of the Si-O-Si bonds~\cite{Shcheblanov2015}. However, it is important to distinguish anharmonic processes (giving rise to irreversible structural changes with energy dissipation - we do not consider here the reversible non-linear mechanical behaviour) and harmonic vibrations in the presence of structural disorder. The last one gives rise to reversible strain heterogeneities that are sufficient to induce "anomalous" (non-Debye) vibrational density of states in the harmonic regime~\cite{Tanguy2002,Schirmacher2013,Mizuno2020, Lerner2021}. The two processes (harmonic and anharmonic) are combined due to the fact that the lowest energy barriers in amorphous materials may be very small~\cite{Hunklinger1995, Mantisi2012, Albaret2017}, and Eshelby inclusions could be traced and anticipated from the reversible soft modes, thanks to extended non-linear analysis~\cite{Maloney2004,Albaret2016,Albaret2017,Tanguy2021,Richard2020}. But the dynamics of these two processes is conceptually very different and affect the thermal properties in different frequency ranges. In the litterature, anharmonicity refers either to the shear transformation zones induced by mechanical instability (whose position can be anticipated by inhomogeneous harmonic properties, like local elastic moduli~\cite{Tsamados2009,Tanguy2010, Mizuno2020b}), either to double well potentials and two-level systems related to the metastability of amorphous materials and the proximity to many different equilibrium states~\cite{HunklingerReview, Angell2005}. In both cases, anharmonicity manifests itself through some amount of temperature dependent dissipated energy having a specific signature in the mechanical constitutive laws~\cite{Albaret2016}, or in the acoustic attenuation~\cite{Mizuno2020}. 

Thermal sensitivity of the mechanical properties of glasses is not obvious, because it combines various origins. Some of them are related to the heterogeneous structure of glasses at the nanometer scale, some other involve anharmonicity and energy barriers distribution. In the high temperature regime for example, even in the supercooled liquid state above the glass transition temperature, viscosity does not follow a simple Arrhenius law~\cite{Angell1991} and this has been interpreted in terms of heterogeneous elastic couplings~\cite{Dyre2015} taking place at high frequency in the supercooled liquid~\cite{Lemaitre2013}, but as well on heterogeneous dynamics~\cite{BerthierReview2011}. Temporal correlations infered from the measurements of the dynamical structure factors in the supercooled liquid (above the glass transition and below the melting temperature) highlighted two kinds of high-temperature relaxation processes: a fast $\beta$-relaxation and a slower streched exponential $\alpha$-relaxation that could be related to a hierarchy or at least to a large distribution of relaxation times~\cite{Karmakar2016}. In the glassy state, below the glass transition temperature, acoustic attenuation follows different frequency laws, interpreted in terms of low to strong acoustic scattering on the elastic heterogeneities~\cite{Tanguy2002,Courtens2003,Schirmacher2013,Mizuno2014,Gelin2016,Damart2017,Luo2022,Szamel2022}, completed by thermal sensitive anharmonic processes that may become dominants in the low frequency regime~\cite{Anderson1972,Phillips1987,Vacher2005,MizunoBarrat2016,Mizuno2020}. The temperature dependence of the thermal conductivity is an example of the anomalous thermal behaviour of glasses compared to crystals. In addition to leading to very low thermal conductivity values, this specific behaviour is remarkable because it is shared by glasses with very diverse compositions, and it gave rise to a lot of different interpretations~\cite{HunklingerReview}. Indeed, in all amorphous materials investigated so far, the low temperature dependence of thermal conductivity shows an unusual power-law with an exponent lower than that of the crystal~\cite{Kittel2004}, followed by a saturation at $10$ to $50 K$, and a then a monotonous increase up to the glass transition temperature~\cite{Zeller1971,Pohl2001,Pohl2002}. In parallel, the low frequency acoustic attenuation in glasses, at a given frequency, shows two peaks in the very low temperature regime (below 50 K)~\cite{HunklingerReview}.  Many different interpretations of this unusual behaviour have already been proposed~\cite{HunklingerReview,Phillips1987,Cahill1987,Allen1993,Vitelli2010}, but up to now they have not been clearly supported by a microscopic interpretation. As already mentioned, heat carriers in dielectric and semi-conductor materials are mainly supported by atomic vibrations, refered to as phonons in crystals~\cite{Kittel2004}. However, how defining phonons in amorphous materials is still a matter of debate~\cite{Heron2009,Maire2017,Beltukov2018}, and the theoretical tools have to be completely renewed, especially to distinguish efficiently harmonic from anharmonic processes.

In this paper, we will first describe the harmonic vibrations in glasses and discuss the role they play in carrying the energy among the samples (part. 2). We will then focus on a comparison between different dissipation processes taking place in glasses (part. 3). In the next part (part. 4), we will discuss the temperature dependence of the thermal conductivity in glasses, after having discussed its atomistic \CORR{foundations}. We will conclude (part. 5) by a discussion and perspectives for future studies.

\section{Vibrational Eigenmodes vs. Phonons in glasses}

\subsection{Eigenvibrations in glasses}

In the harmonic approximation, the vibrations, in $3D$ glasses containing $N$ atoms, are obtained by \CORR{diagonalizing} the $3N\times3N$ Dynamical Matrix $\{ D_{i j}^{\alpha\beta} \}$ with 
\begin{equation}
D_{i j}^{\alpha\beta}=\frac{1}{\sqrt{m_i m_j}}\frac{\partial E}{\partial r_i^\alpha\partial r_j^\beta}
\end{equation} 
defined from the second order derivatives of the interatomic potential energy $E$  as a function of the positions $\vec{r}_i$ and $\vec{r}_j$ of atoms $i$ and $j$ with mass $m_i$ and $m_j$ respectively, $\alpha$ and $\beta$ being the components of the vectors. The eigenvectors (also called {\it Normal modes}) $\vec{e}_n=\{e_i^\alpha(n)\}$ of the Dynamical Matrix are orthogonal in the harmonic limit, thus acting as a well defined basis to perform the modal decomposition of any given motion. In the harmonic approximation, the eigenvectors do not interact with each other. According to the equations of motion, the atomic displacements related to mode $n$ are $u_i^\alpha(n)=e_i^\alpha(n)/\sqrt{m_i}$, and the eigenfrequency $\omega_n$ of mode $n$ is given by the square-root of the corresponding eigenvalue of the Dynamical Matrix. This one is positive, since the interactions are computed around a stable mechanical equilibrium. The eigenmodes are thus obtained by \CORR{diagonalizing} a very large $3N\times3N$ positive-definite matrix and result, by construction, of collective motion. 

In crystals, these eigenmodes are called "phonons". They all are identified by a wave vector $\vec{q}$ related to their periodicity along the crystal, and given by $2\pi/\lambda_\alpha(n)$ in each direction $\alpha$, where $\lambda_\alpha(n)$ is the wavelength of the mode. There are always three acoustic modes in crystals (two transverse and one longitudinal) and $3R-3$ optical modes (with $R$ the number of atoms in the primitive cell). The wave vector $\vec{q}$ corresponds to the momentum (related to the kinetic energy $E_c=\hbar^2q^2/2m$) in the quantum description of motion. The related quantum of energy is given by $\hbar\omega_n$. At wavelengths in the nanometer-subnanometer range, the phonon frequency is typically of the order of a few terahertz ($THz$), corresponding to the "hypersonic regime" with phonon energies of a few meV ($1 THz \approx 4.18 meV \approx 33.7 cm^{-1}$). At macroscopic wavelengths, acoustic phonons are also identified with the elastic waves, as described by the continuum elasticity theory. In crystals, phonons are thus responsible for \CORR{sound propagation}, but also for heat (or energy) \CORR{transport}.

In disordered materials, the definition of phonons is more intricate. In this case indeed, the normal modes are in general not characterized by a single wave vector. There are thus different perspectives: either considering phonons as quasi-particles with wave-like characteristics that can be strongly scattered in the material~\cite{Hardy1963}, either relating phonons to vibrational eigenmodes. In the last case, many of them cannot be characterized by a wavelength~\cite{Lv2016}. In amorphous materials, normal modes are grouped indeed into three categories~\cite{Allen1999}: the {\it propagons} (including harmonic soft modes) wich have a dominant wavelength, the {\it diffusons} which look like a random superposition of plane waves, and the {\it locons} with a multifractal shape similar to Anderson’s localized modes induced by disorder~\cite{Beltukov2018,Anderson1958,Castellani1986a,Hu2008,Beltukov2017}. \CORR{ In general, propagons dominate the low frequency regime, but propagons with longitudinal dominant polarization may also appear at frequencies larger than that of some diffusons, for example in cases similar to amorphous silicon (see figure~\ref{fig:VDOS}-top), where there is a large frequency gap between dominantly transverse and dominantly longitudinal vibrations~\cite{Beltukov2016}. Locons form in general at a frequency where the vibrational density of states decreases abruptly (transition to a gap), at the so-called {\it mobility edge}.} Typical examples of these modes obtained in a numerical description of amorphous silica are shown in figure 1. As shown in the figure and confirmed in a deepened numerical analysis~\cite{Beltukov2016,Tanguy2002}, {\it diffusons} and {\it locons} cannot be characterized by a single wavelength but by a large set of wavelengths.  {\it The two definitions of phonons (with or without wavevector) are thus not equivalent}. Actually, the reference to experimental wave-like excitation, for example using X-Ray scattering, is in favor of the definition of phonons involving a wavevector. It can then be noticed that phonons defined in this way naturally interact with each other in the harmonic regime, even at very low temperature, since they are not normal modes. Most confusions due to references to "hybridization" or to "anhamornic coupling between phonons" results certainly in glasses from the fact that harmonic vibrations do not refer to phonons ideally defined only in perfect crystals, and in the same way from the fact that the ideal phonons appears as "coupled" on nanometric wavelength scales, since they are certainly not independent normal modes, in this range. In the following, we will thus identify phonons as wave-packets with well defined central frequency and wave fronts, that corresponds to classical descriptions of quantum quasi-particles with energy $\hbar\omega$ and wavevector $\vec{q}$.

\begin{figure}[tbp]
\includegraphics[width=0.33\linewidth]{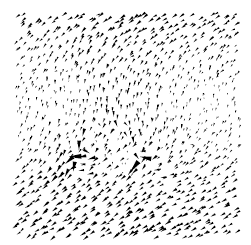}
\includegraphics[width=0.47\linewidth]{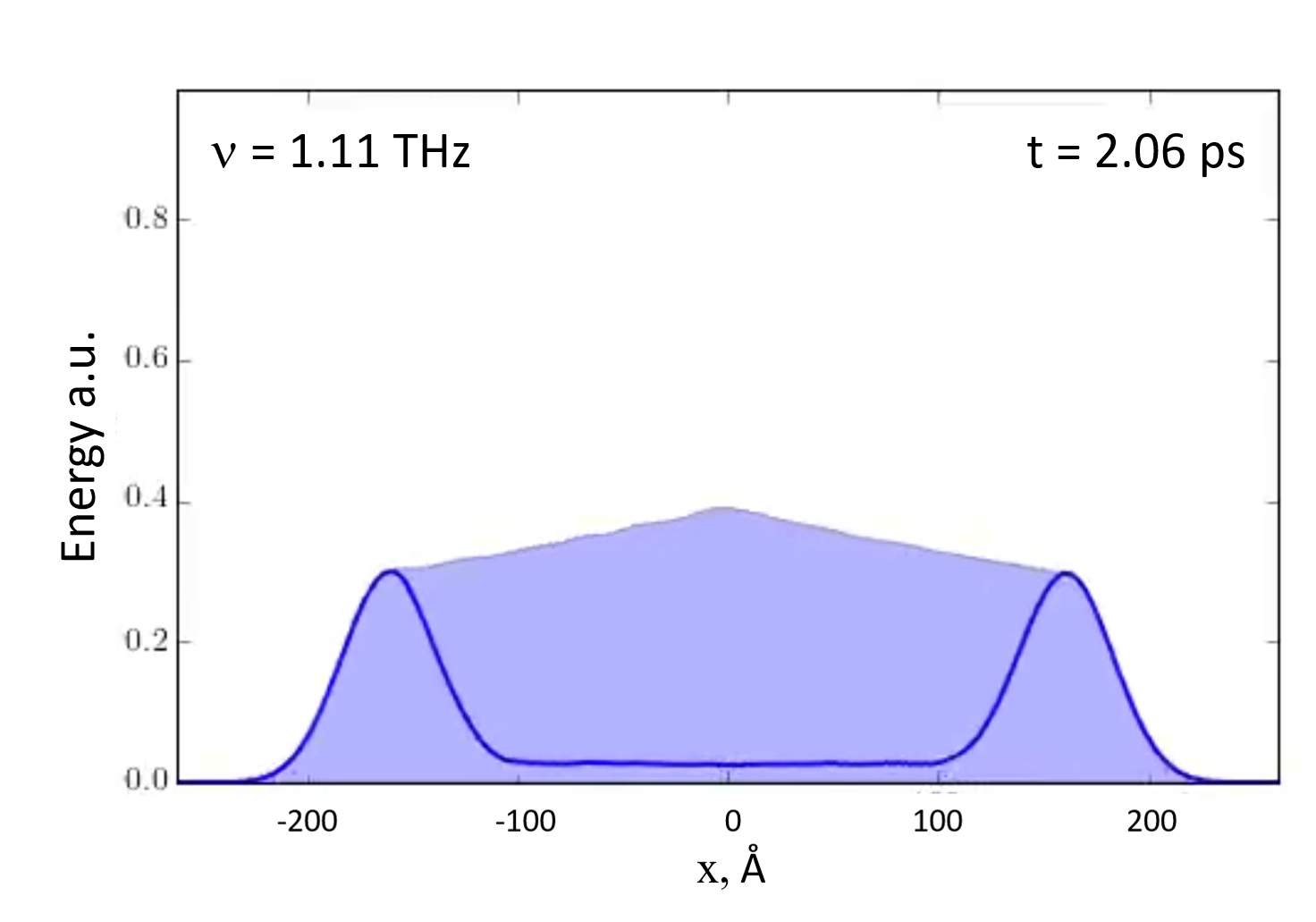}
\includegraphics[width=0.33\linewidth]{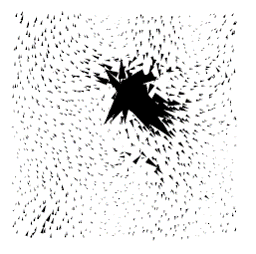}
\includegraphics[width=0.47\linewidth]{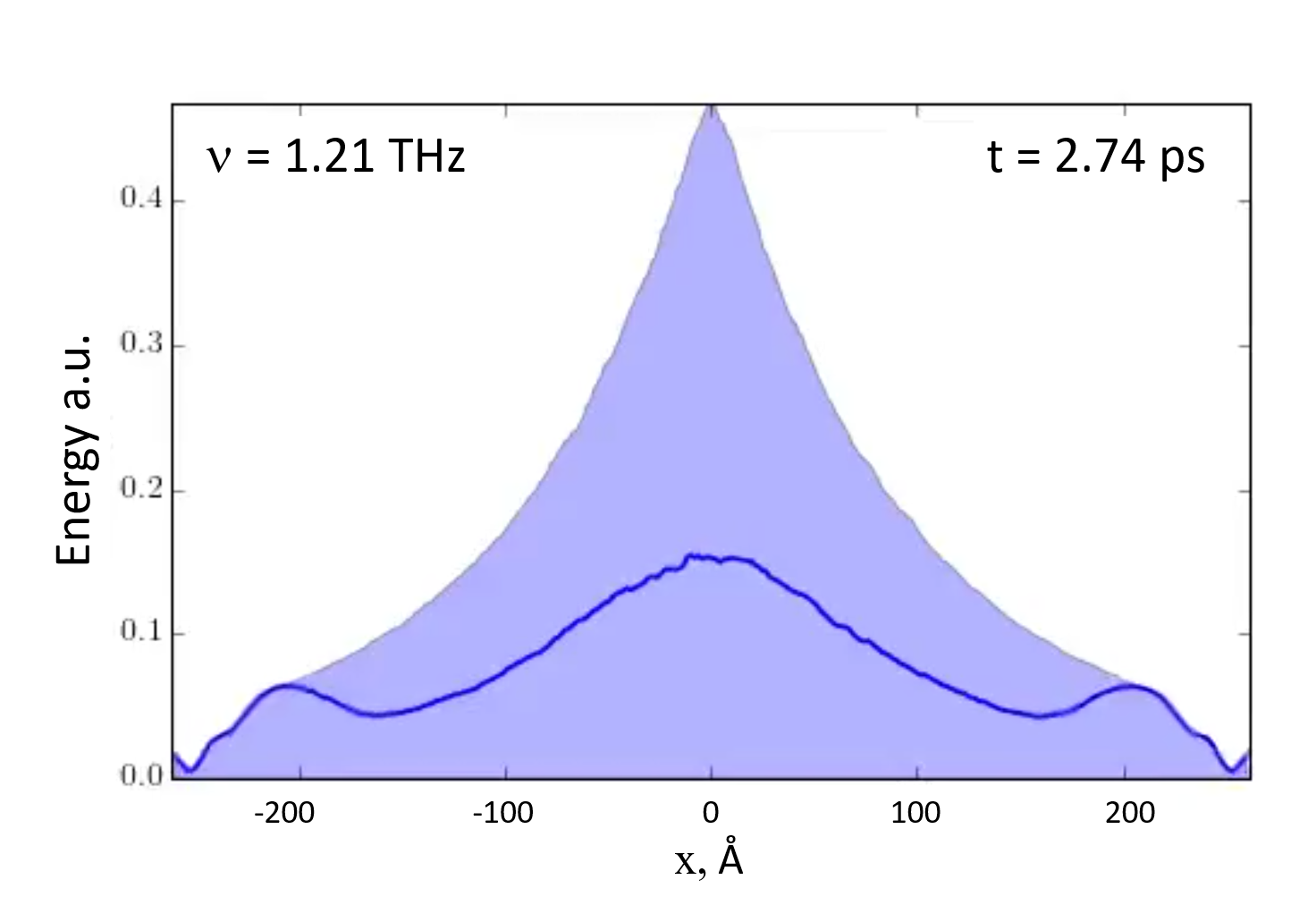}
\includegraphics[width=0.33\linewidth]{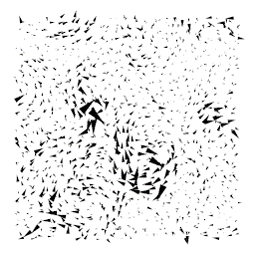}
\includegraphics[width=0.47\linewidth]{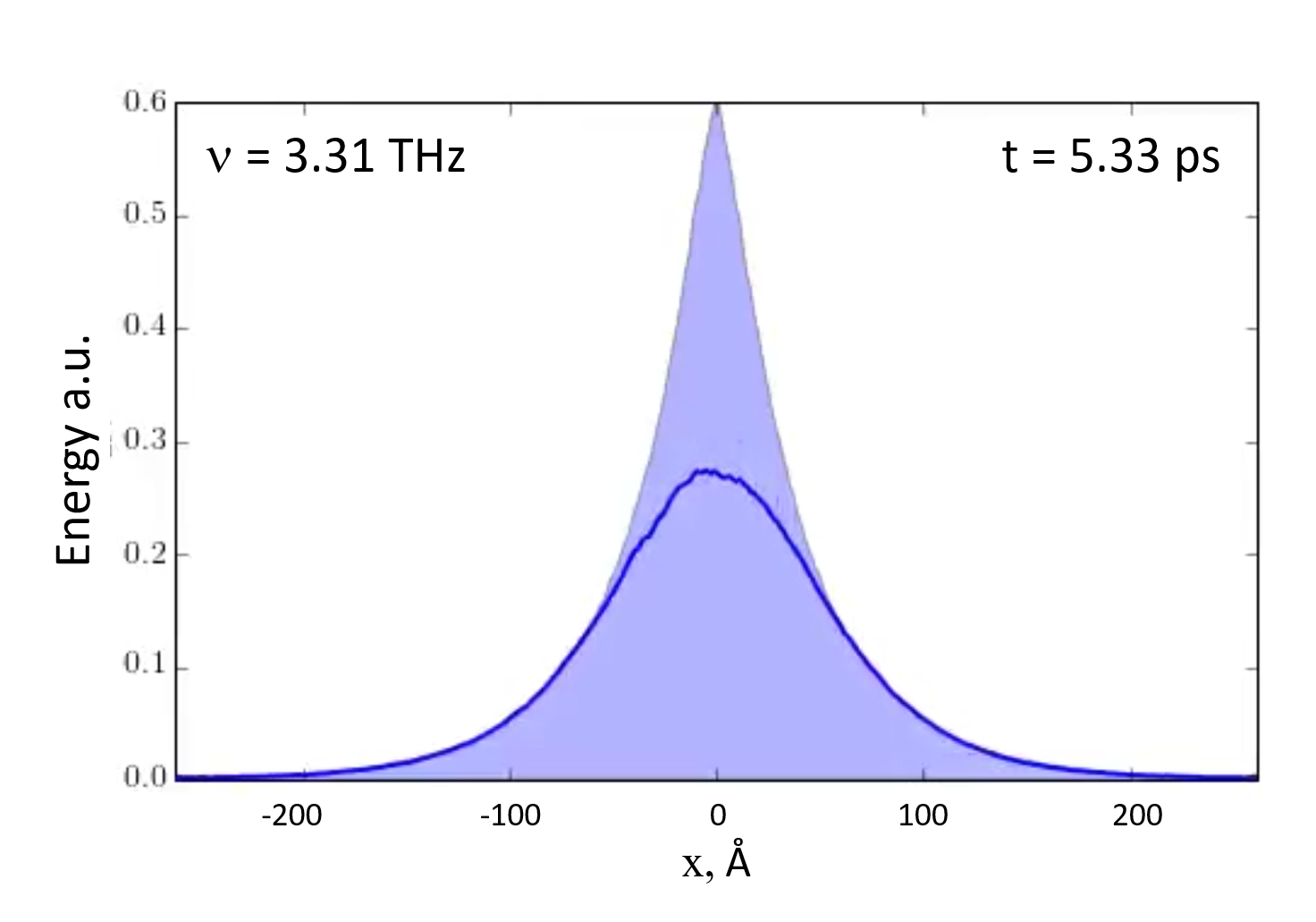}
\includegraphics[width=0.33\linewidth]{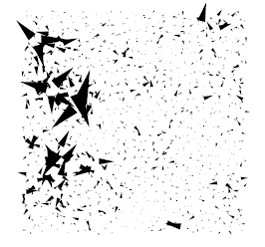}
\includegraphics[width=0.47\linewidth]{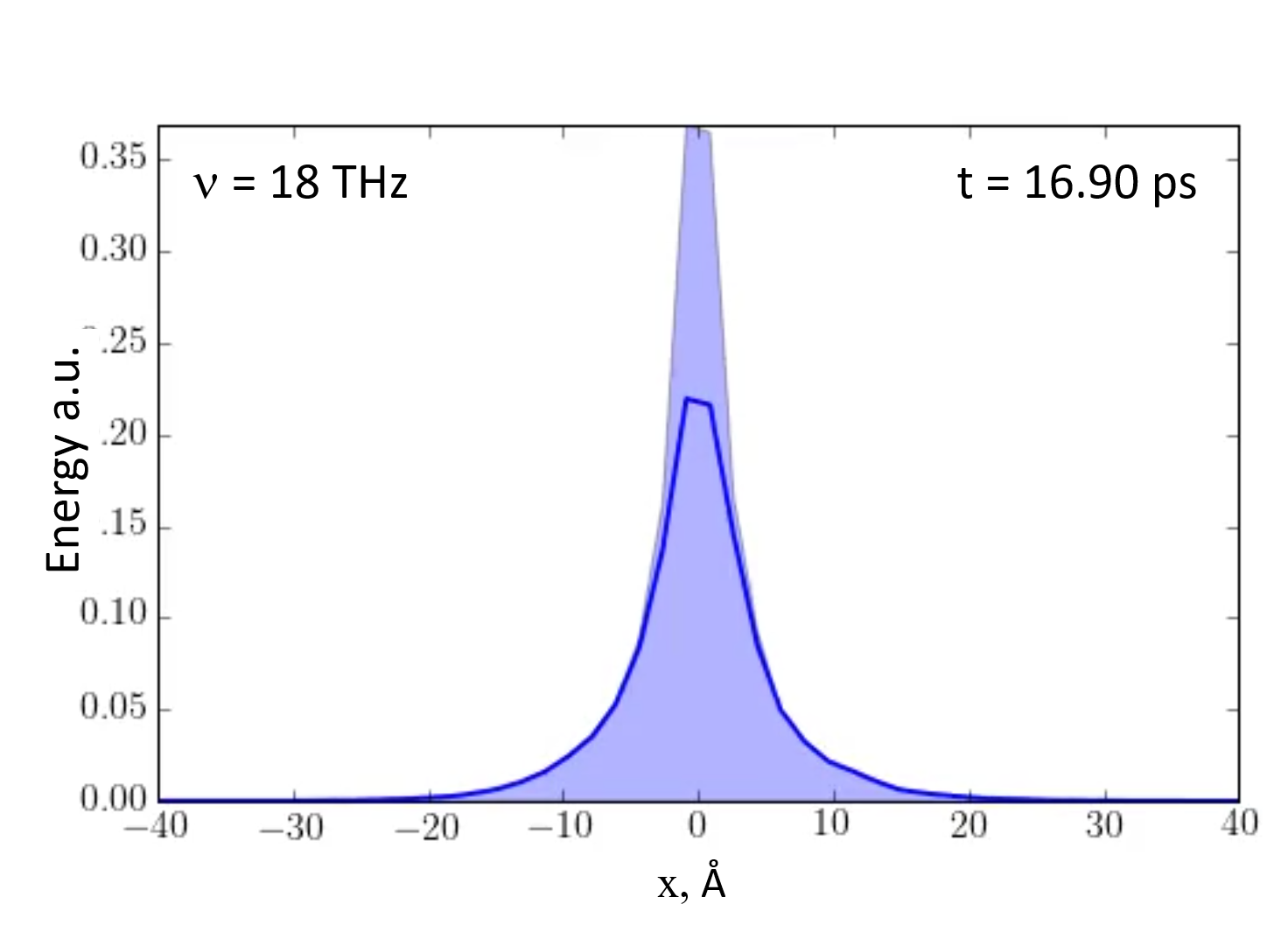}
 \caption{Left column: Normal modes in a numerical model of a silica glass~\cite{Mantisi2012}. The arrows are the instantaneous displacement vectors of the atom located at the basis of the arrow. From top to bottom: propagon, soft mode, diffuson, locon. 
Right column: Enveloppe of wave packets propagation for different excitation frequencies~\cite{Beltukov2018}. \CORR{The blue curve is the instantaneous and smoothed kinetic energy at the time indicated in the inset as a function of the position x to the central excitation. The colored region corresponds to the enveloppe of the kinetic energies for the previous times.} From top to bottom: exponential attenuation of the enveloppe in the frequency range of propagons, mixed regime, gaussian enveloppe with diffusive transport in the frequency range of diffusons, and localized regime (no energy transfer). }
 \label{fig:VibModesEnv}
 \end{figure}
                                                                                                                                                                                                                                                                                                                                                                                                                                                                                                                                                                                                                                                                                                                                                                                                                                                                                                                                                                                                                                                                                                                                                                                                                                                                                                                                                                                                            
Concerning the energy transportation by wave-packets, it has been confirmed recently~\cite{Beltukov2018}, that the enveloppe of a wave-packets excitation in the frequency range of {\it propagons} is smoothly attenuated, following a spatial exponential decay that allows defining properly a mean-free path as a function of the frequency (see figure 1-right first image from the top). The exponential decay is due to the small amount of energy going back due to the local impedance mismatch resulting from the structural disorder. In the frequency range of {\it diffusons}, the enveloppe of a wave-packets excitation is enlarged and attenuated diffusively, thus preventing the identification of a group velocity, but allowing instead the identification of a {\it diffusivity coefficient} (see figure 1-right third image from the top). It has to be noted that the transition from {\it propagons} to {\it diffusons} involves a mixed regime with coexistence of a ballistic attenuation of the wavefront (thus with a characteristic mean-free path) and a diffusive part contributing to the aftershocks already evidenced in geophysics~\cite{Ishimaru1978}. In the frequency range of {\it thermal phonons} that will be discussed in part 4, the diffusive part corresponds strictly to diffusive heat transfer. The relative amplitude of the diffusive part with respect to the propagative one increases  with the frequency, or with the distance to the excitation since the diffusive contribution is less attenuated with time. The diffusive part thus overcomes progressively the ballistic part of the excitation. Finally, in the tiny frequency range of {\it locons}, in general in the frequency ranges corresponding to energy gaps in the crystals, wavepackets do not propagate anymore, thus preventing the transportation of energy. Only the {\it{locons}} are relevant for the localization, as predicted by Anderson~\cite{Anderson1958,Castellani1986a, Mirlin2000, Mirlin2008}. \CORR{Their shape has the multifractal characteristics of the wavefunctions at the mobility edge of the Anderson's transition~\cite{Beltukov2016,Mirlin2008},} and the numerical simulations confirm that they keep the energy where it is deposited, thus preventing completely its propagation along the system~\cite{Beltukov2018}. Note that the situation is not the same in the case of low frequency soft modes, because soft modes do not prevent the propagation of energy by the large wavelength elastic component of the vibration, except in the extreme case when the system approaches its instability threshold and behaves strongly anharmonicly. This is for example the case in granular solids close to the jamming transition.

\subsection{Vibrational Density of States}

\begin{figure}[tbp]
\includegraphics[width=0.65\linewidth]{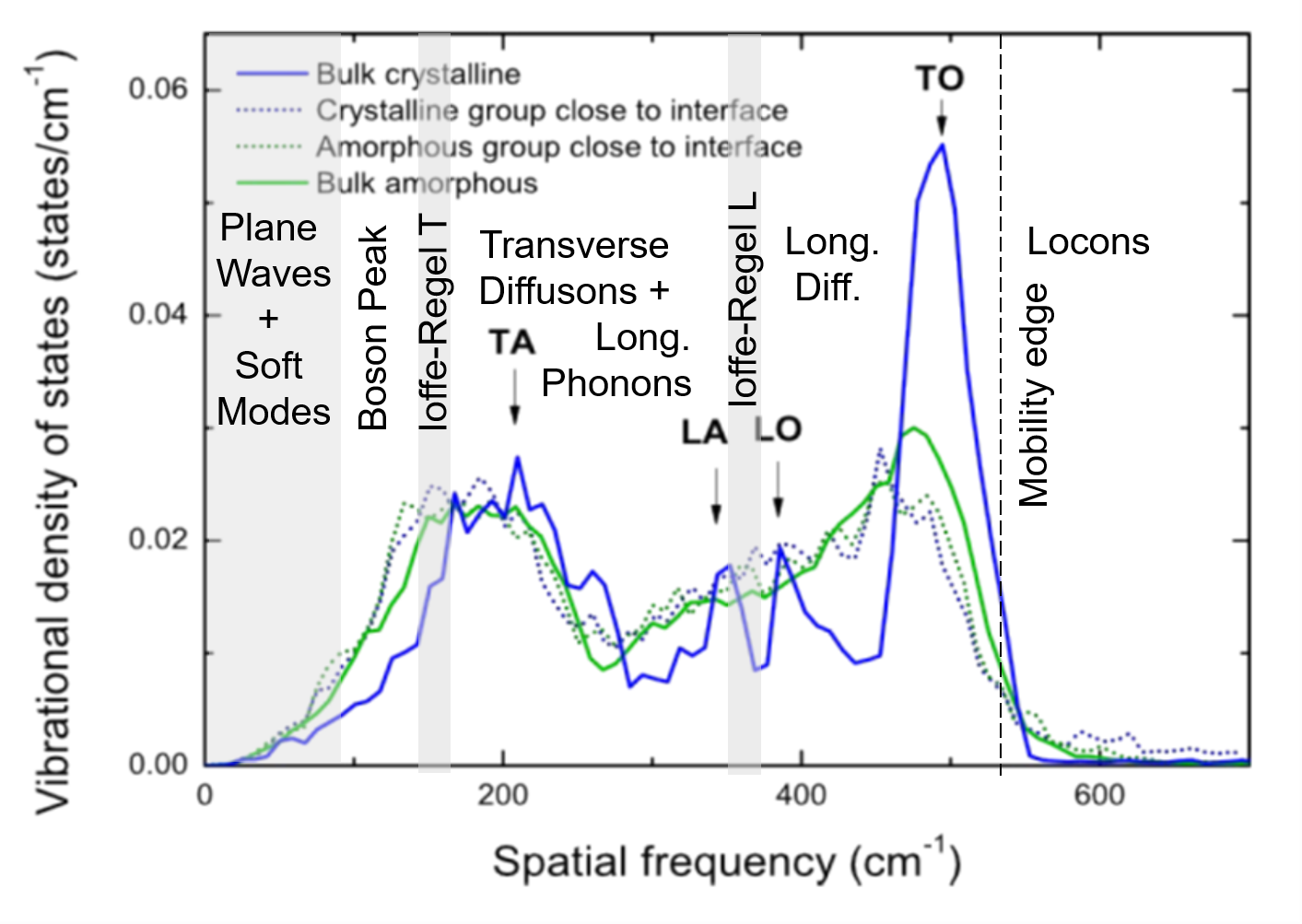}
\includegraphics[width=0.65\linewidth]{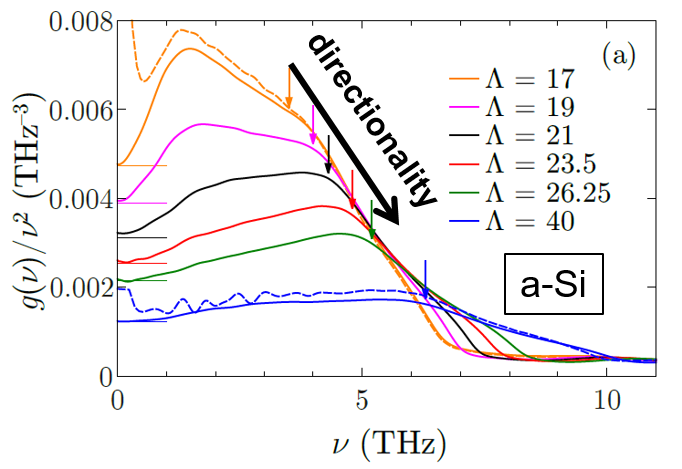}
\includegraphics[width=0.65\linewidth]{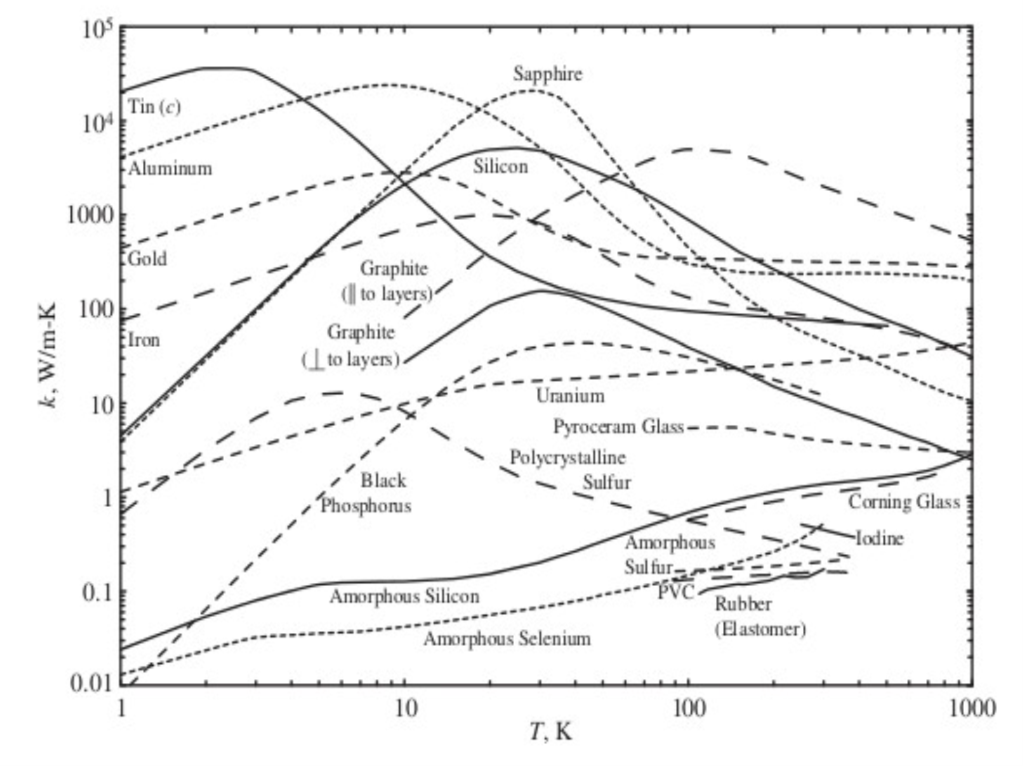}
 \caption{Top: Effect of amorphization on the vibrational density of states of silicon samples obtained with Stillinger-Weber (SW) potential~\cite{Stillinger1985}. \CORR{ In the different regimes, the related types of vibrations for the amorphous samples is specified in accordance  with~\cite{Beltukov2016}.} Middle: role on the Boson Peak of the relative weight of the three-body term in the SW potential, for silicon-like amorphous samples~\cite{Beltukov2016}. Bottom: Thermal conductivity as a function of themperature in crystals and in amorphous materials with various compositions~\cite{Touloukian1966}}
 \label{fig:VDOS}
 \end{figure}

Another characteristics of amorphous materials having consequences on their thermal properties is the "anomalous" shape of their density of vibrational states (VDOS), as compared to the crystals. As seen in figure 2-Top, the amorphisation of the structure induces an enlargment of the different peaks in the vibrational density of states, and especially an increase in the density of low frequency modes. The ratio between the low frequency VDOS and the Debye prediction valid in crystals (Debye VDOS $g(\omega)\propto\omega^2$), gives rise to a "Boson Peak" in the low frequency regime ($\approx 0.5 THz$) having a marqued effect on the heat capacity for temperatures above 20 K (when frequencies at Boson Peak start to matter - see figure 3). This Boson peak deserves different explanations. It is located at a characteristic frequency close to the Ioffe-Regel frequency ($\omega_{IR}$ where the mean-free path of phonons becomes smaller than their wavelength) and the related characteristic length $\xi=c_T/\omega_{IR}$ changes with the nature of the bonds inside the glass, jumping from $5$ to $25 \AA$ when the bending rigidity is decreased~\cite{Beltukov2016}. At smaller frequencies, or more precisely for wavelengths above this mesoscopic lengthscale, stable glasses are mechanically homogeneous and isotropic~\cite{Tsamados2009}, and the Boson Peak is located at a cross-over frequency separating the low scattering regime from the strong scattering regime of acoustic waves on elastic heterogeneities~\cite{Beltukov2016,Tanguy2010}. \CORR{In terms of eigenmodes, the Boson Peak is located at the transition between phonons and propagons. This means that the} Boson Peak is already visible in the harmonic limit~\cite{Beltukov2016, Tanaka2022}, thus the existence of such a peak does need neither anharmonicity neither stability loss, but additionnal anharmonic effects \CORR{can increase its amplitude even more}~\cite{Kojima1996}. One possible explanation is to relate it to elastic heterogeneities{\cite{Schirmacher1998a, Tanguy2002, Leonforte2006, Schirmacher2007, Mizuno2014,Mizuno2019}. The decay of the effective sound velocity as compared to crystals results directly from the large distribution of strain heterogeneities~\cite{Torquato2001,Tanguy2002}  at the scale contributing to the vibrations. The consequence is the decay of the related eigenfrequencies. In the harmonic limit and when the bending rigidity of bonds is decayed, its maximum amplitude is strongly amplified and shifted to lower frequencies (see Fig.2-Middle), implying that larger scales are involved. This peak can be used as a suitable size independent indicator for the intrinsic nanometer lengthscales characteristic $\xi$ of collective mechanical heterogeneities taking place in amorphous solids~\cite{Tanguy2002, Leonforte2006, Tanguy2010, Beltukov2016}. Interestingly, such an intrinsic lengthscale is not directly related to the size of a structural defect, but results from a collective motion induced by a combination of elastic bonding and geometrical frustration.

\begin{figure}[tbp]
\includegraphics[width=0.5\linewidth]{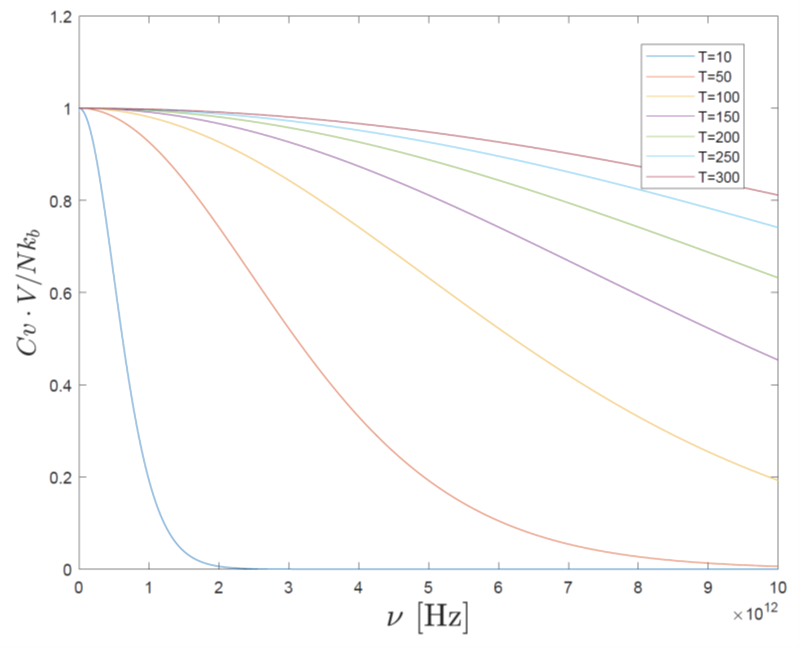}
\includegraphics[width=0.45\linewidth]{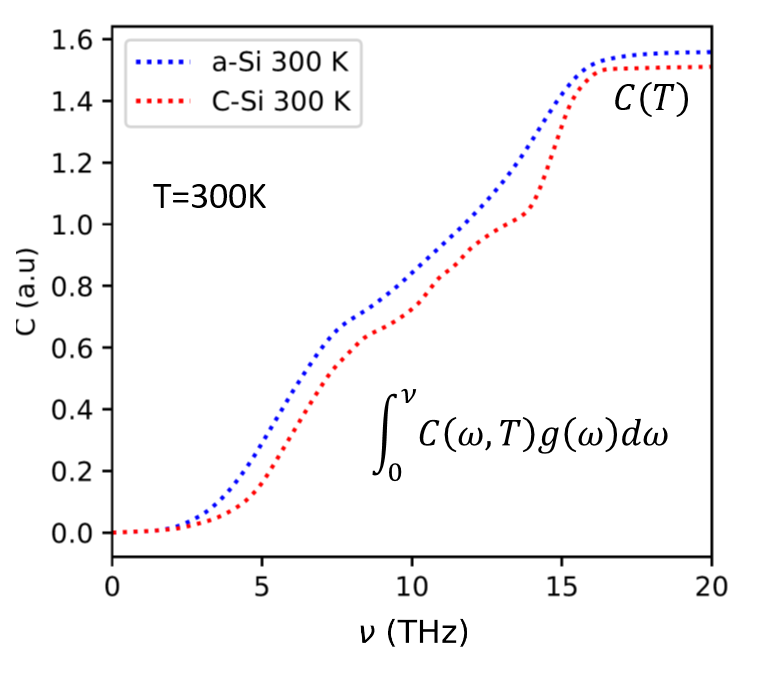}
 \caption{Debye Model of the Heat capacity: (left) frequency dependence of the modal contribution to the Debye heat capacity at different temperatures. (right) cumulated heat capacity for a model of amorphous (blue) and crystalline (red) silicon at $T=300 K$. The gap between the two curves hangs from the frequency of the boson peak,  and is attenuated when taking into account high density high frequency optical modes (that are involved at sufficiently high temperature) }
 \label{fig:Cv-Si}
 \end{figure}
 
Various articles focused as well on the density of "soft spots" in the very low frequency regime, maybe taking place in the gap between successive plane waves~\cite{Lerner2021}. In all the amorphous systems we have studied~\cite{Tanguy2010,Mantisi2012,Molnar2016,Molnar2017,Beltukov2016}, we have observed "soft spots" whose frequency decreases when the system becomes closer to a mechanical instability as already observed by A. Lemaitre et al.~\cite{Maloney2004, Tanguy2010, Rodney2011}. But in our case, these vibrations were always coupled to elastic plane waves~\cite{Beltukov2016}. The $\omega^4$ dependence of the density of soft spots mentioned in~\cite{Lerner2021} can be reproduced only when averaging over a very large number of samples, thus corresponding indeed to the distribution of the lowest elastic moduli at the soft spot (since it is related to its eigenfrequency). In the frequency range of {\it soft spots}, we want to stress here the elasticity plays certainly its usual role for energy transfer of wavepackets excitations, as long as the systems is mechanically stable, and since it is homogeneous on the large scale corresponding to the wavelength of the incident excitation~\cite{Tsamados2009}. It is not the case in the frequency range of {\it locons} where elastic excitations are strongly scattered, preventing elasticity to play its long-range coupling role anymore. In the frequency range of locons, the energy stays at the place where it has been deposited, it is decoupled from the rest of the solid, and is not transfered across it (localization of energy). Note that, contrary to usual meaning, the frequency range of locons is not the low frequency regime, but it is system dependent.  In the very low frequency regime, anharmonicity can play an important role, due to the possibility to generate a mechanical instability having a signature at zero frequency. When the energy barrier becomes smaller, the system becomes easily able to transit from one mechanical equilibrium to another. This situation is related to the progressive decay of a positive eigenfrequency upon a destabilizing mechanical load. But the low frequency modes are also the modes having the most important contribution to the heat capacity in the low temperature regime (see figure 3-a). In the litterature, these modes also refer to the two-level systems~\cite{Phillips1987,Anderson1972} at very low temperature (below $20 K$). Two-level systems are related to a "resonant absorption" of energy induced by a tunneling transition between two different energy states thanks to phononic interactions. At slightly higher temperatures ($\approx 20-50 K$) they refer to thermally activated relaxations in a double-well potential~\cite{HunklingerReview}. Both descriptions are related to anharmonic transitions from one equilibrium to another, with different kinds of activations, and different amplitudes of energy barriers (higher in the double-well potential model than in the two-level systems) as discussed in Ref.~\cite{HunklingerReview}. Up to now however, no definite experimental proof of such models nor identification of the related equilibrium positions have been given at the atomic scale, but they clearly allow an efficient formal description of the low temperature ultrasonic and thermal properties of glasses. In the next part of this paper, we will discuss these models in the context of the temperature dependence of the thermal conductivity.

\section{A variety of dissipation processes in glasses}

In glasses, dissipative processes are unlikely related to well identified defects. New kinds of dissipative processes occur, that can be independent on the specific chemistry of the material. Let's first consider the energy transfer supported by a wave packet in the purely harmonic regime.

\subsection{Harmonic processes of acoustic attenuation}

\begin{figure}[tbp]
\includegraphics[width=0.7\linewidth]{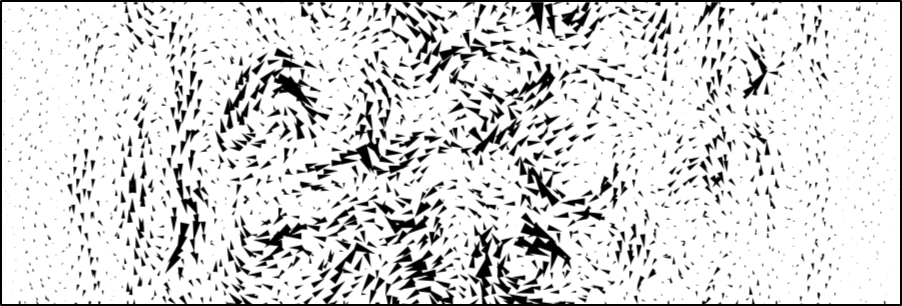}
\includegraphics[width=0.35\linewidth]{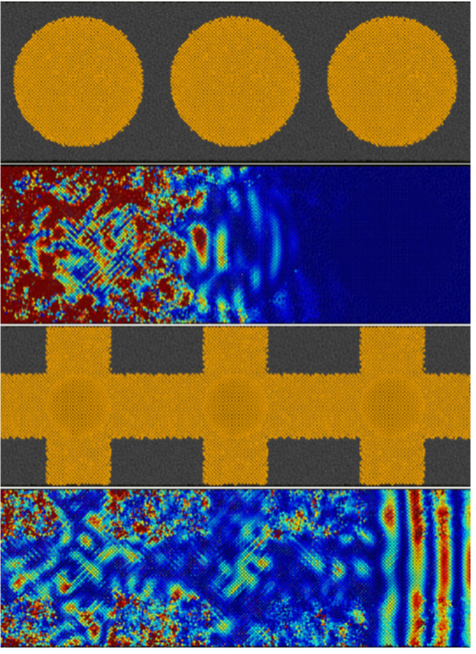}
\includegraphics[width=0.35\linewidth]{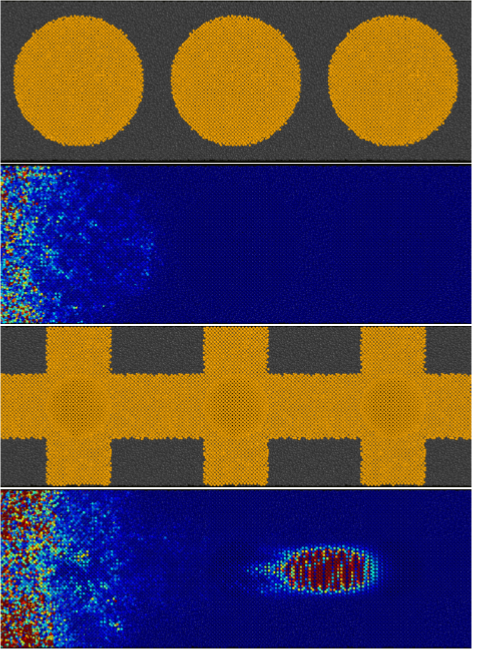}
\caption{Top: Wave packet excitation in an amorphous Si sample via Molecular Dynamics simulations, in the mixed regime ($t=1.5 ps, \omega = 2.2 THz$), black arrows indicate the instantaneous atomic displacements~\cite{Beltukov2018}. Middle: Wavepacket excitation in silicon nanowires with amorphous matrix for $\omega=2THz$ (left) and $\omega=11THz$ (right). Bottom: Idem in silicon whiskers with amorphous matrix for $\omega=2THz$ (left) and $\omega=11THz$ (right), colors indicate the local kinetic energy. }
 \label{fig:WavePack}
 \end{figure}

\begin{figure}[tbp]
\includegraphics[width=0.4\linewidth]{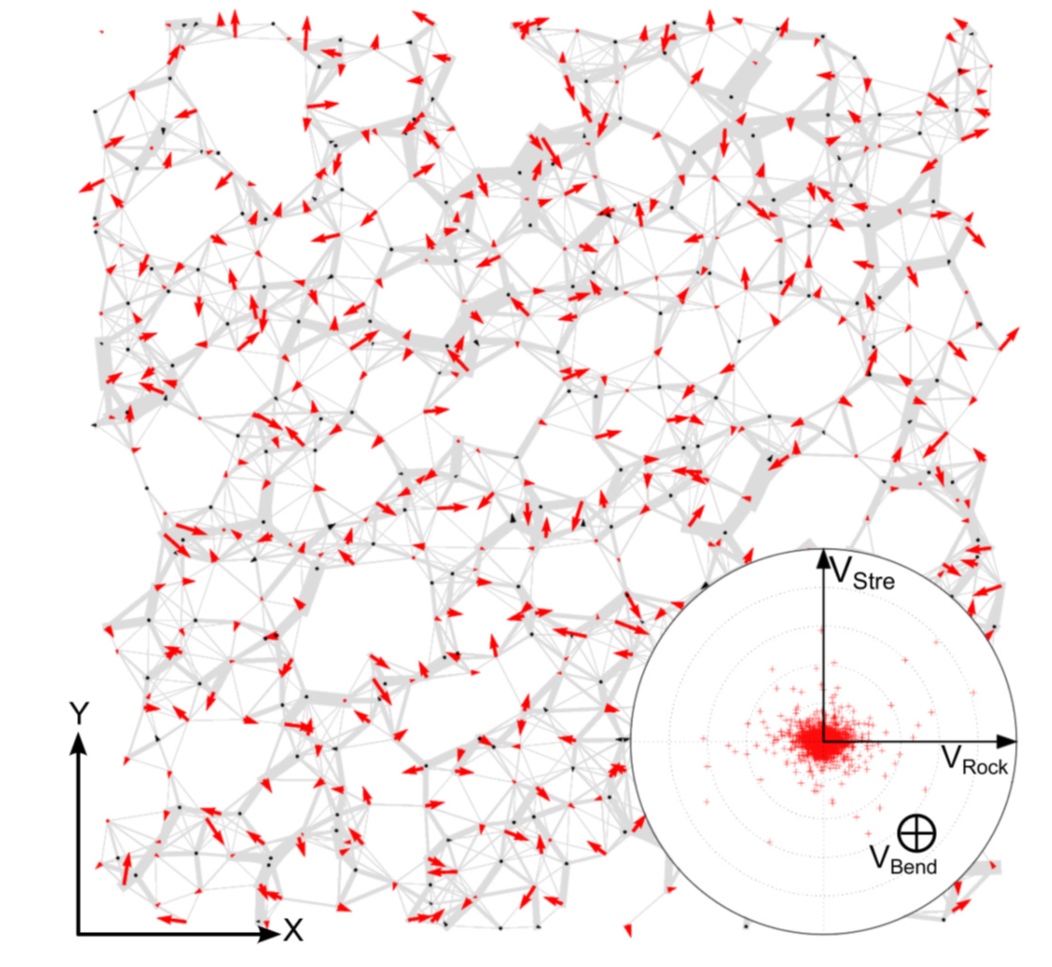}
\includegraphics[width=0.55\linewidth]{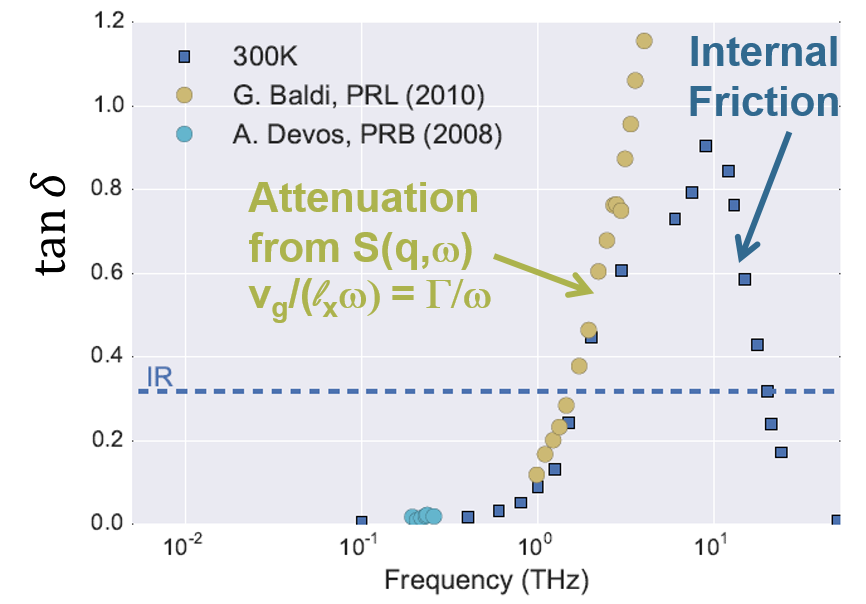}
 \caption{Illustrations of harmonic dissipation mechanisms: (left) structural heterogeneous force links (see~\cite{Damart2017} for more details) in a numerical model of $SiO_2$ submitted to oscillatory compression, represented in two-dimensional projection for an $8\,\AA$ slab in a $SiO_2$ sample. Black arrows are for $Si$ atoms, red arrows for oxygen atoms. (right) internal friction vs. longitudinal acoustic attenuation in the harmonic model of dissipation in the same $SiO_2$ sample~\cite{Damart2017}.}
 \label{fig:Dissip}
 \end{figure}

In the harmonic regime, in the absence of microscopic dissipation, the global energy is conserved. But due to impedance mismatch resulting from elastic heterogeneities at the atomic level, part of the moving energy is repelled in the direction opposite to the direction of propagation. In the ballistic regime at low frequencies (frequency range of {\it propagons} - figure 1-right, first image from the top), due to the small amount of energy repelled at each step $$dE=-aEdx,$$ the enveloppe of energy is attenuated following an exponential Beer-Lambert Law $$E=E(0)\exp{\left(-ax\right)}$$ This allows defining properly a mean-free path $$\ell_{MFP}=\frac{1}{a}$$ that depends on the frequency $\omega$ of the excitation (that is of the exciting wave-packet). From this mean-free path, an acoustic attenuation $\Gamma$ may be defined in analogy with damped harmonic oscillators $\Gamma=v_g/\ell_{MFP}$ with $v_g$ the group velocity of that wave, or equivalently $\Gamma=1/\tau$ with $\tau$ the effective relaxation time. There are different explanations for the origin of the temperature and frequency dependence of $\ell_{MFP}$. We will review some of them here, starting from those involving only harmonic couplings between atoms.

When \CORR{weak} scattering is the source of energy deviation~\cite{Gelin2016,Szamel2022,Mizuno2020,Ishimaru1978}, then $$\ell_{MFP}\propto\omega^{-4}$$ The energy repelled is mainly contained in rotational non-affine motion~\cite{Tanguy2002,Beltukov2018,Szamel2022,Tanaka2022} visible in figure 4-a, that will generate the after-shock in the mixed regime - where part of the energy is kept in plane waves, and a similar part is stored into rotational motion (as illustrated in figure 1-right second line for the enveloppe of energy). Above the Ioffe-Regel frequency, the mean-free path of the wavepackets becomes smaller than their wavelength: this corresponds to the frequency range of {\it diffusons}. In this range, wave-fronts are destroyed as soon as they are projected on the eigenmodes of the system, and the plane wave looses its coherence. The resulting propagation of energy follows a diffusive equation due to concurrent directions of propagation~\cite{Beltukov2018}. The diffusive motion of the wave-packets that represent units of energy carriers, or phonons, must of course be distinguished from the usual dynamical motion of diffusons, that correspond to the harmonic relaxation-free eigenvibrations. This distinction has been clarified in our Ref.~\cite{Beltukov2018} as it was a little confused in earlier litterature~\cite{Allen1993,Larkin2014}. In this diffusive regime, it is impossible to identify a group velocity for phonons, and the notion of mean-free path is meaningless as clearly seen in figure 1-right (third image from the top). It is only possible to identify a diffusivity coefficient for the kinetic energy, strictly analogous to thermal diffusivity as will be discussed later. Finally, in the frequency ranges of {\it{locons}} (figure 1-right last image from the top), the energy is kept where it has been deposited, phonons are pinned in the material and completely decoupled from the rest of the solid. In this frequency range, the system is an acoustic insulator~\cite{Beltukov2018,Hu2008,Castellani1986a,Beltukov2017}.   

Another approach to dissipation in the amorphous materials is related to the application of mechanical oscillatory excitation, in analogy with rheological analysis. The global dissipation is then highlighted due to the projection of the atomic displacements on the eigenmodes (modal decomposition), as soon as some microscopic dissipative process is taken into account~\cite{Damart2017}. The microscopic process is always present in solids: it may be due to local electronic excitation/desexcitation and Fermi rule~\cite{Sokoloff1995}, to local heating or noise emission~\cite{Persson1985}, or to the intrinsic fluctuation-dissipation theorem for isothermal loading~\cite{Kubo1966}. The characteristic relaxation time for these excitations is always of the order of $10^{-13} s$, that corresponds as well to the inverse maximum bonds vibration frequency. In the case of oscillatory imposed relative volume change $\epsilon$, the equations of motion written in terms of the non-affine displacements on the atom $i$ give
\begin{equation}
	m_i\ddot{x}_i^{\alpha}=-\left(\sum_{j\beta}\sqrt{m_im_j}D_{ij}^{\alpha \beta}R_{ij}^{\beta}\right)\epsilon-\sum_{j\beta}\sqrt{m_im_j}D_{ij}^{\alpha \beta}x_{j}^{\beta}-m_i\gamma\dot{x}_i^{\alpha}+F_{th}
\label{eq:motion}
\end{equation}
with $D_{ij}^{\alpha \beta}$  the matrix of the second order derivatives of the potential energy, $m_i$ the mass, $R_{ij}^\beta$ the equilibrium separation between atoms $i$ and $j$ in direction $\beta$ in the underformed initial cell, $F_{th}$ the thermal noise and $\gamma$ the massic viscosity, or Langevin friction.  This equation may be solved using the projection coefficients $s_n$ on the eigenmodes $\{e_i^\alpha(n)\}$. Using 
\begin{equation}
	\sqrt{m_j} x_j^{\beta} = \sum_n s_n e_j^{\beta}(n)
\label{eq:proj}
\end{equation}
the equations of motion~(\ref{eq:motion}) are rewritten~\cite{Damart2017}
\begin{equation}
	\ddot{s}_n=C_n \epsilon- \omega_n^2s_n-\gamma\dot{s}_n+F_n
\end{equation}
with
\begin{equation}
	C_n = \sum_{i\alpha j\beta}\sqrt{m_im_j}D_{ij}^{\alpha \beta}R_{ij}^{\alpha}\frac{e_j^{\beta}(n)}{\sqrt{m_j}} 
    \label{eq:CmAppen}
\end{equation}
is the pressure induced by the forces supported by the bonds upon incrementing the atomic displacements along the eigenmode $n$ (see Fig.5-left). That quantity $C_n$ is \CORR{unlikely to be predicted} in amorphous materials, but it is zero in case of a cubic crystal with only two-body forces. It controls however the internal friction quantifying the amount of dissipated energy related to the amount of energy stored during one temporal period $T_p$
\begin{eqnarray}
\tan\delta&=&\frac{1}{2\omega}\frac{\int_0^{T_p/2}<P>(t)\dot{\epsilon}(t)\,dt}{\int_0^{T_p/2}<P>(t){\epsilon}(t)\,dt}\nonumber\\
&=&\frac{\sum_n C^2_n \frac{\omega\gamma}{(\omega_n^2-\omega^2)^2+(\gamma\omega)^2}}{\frac{9V_0}{2}K^\infty - \sum_n C^2_n \frac{\omega_n^2-\omega^2}{(\omega_n^2-\omega^2)^2+(\gamma\omega)^2}}\propto\frac{\Gamma}{\omega}=\frac{c_L}{\omega\ell_{MFP}}
    \label{eq:IntFri}
\end{eqnarray}
where $\Gamma$ is also the acoustic attenuation of longitudinal waves in the ballistic regime (see Fig.6-b) and $c_L$ the longitudinal sound velocity~\cite{Damart2017}. \CORR{The local Langevin damping $\gamma$ contributes here mainly to the smooting the data as function of the frequency. The mentioned relation of the internal friction to the acoustic attenuation $\Gamma$ in~Eq.(\ref{eq:IntFri}) results directly from the resolution of continuous waves equations in the Fourier space, using the real and imaginary parts of the frequency dependent shear moduli. The imaginary part yields to spatial attenuation on a distance $c_L/\Gamma=2c_L/(\omega.\tan\delta)$ as shown in textbooks~\cite{Sorbonne}. } The calculation made in~\cite{Damart2017} shows that the non regular shape of vibration modes described in Part.2 (Fig.1) coupled to usual atomic dissipative processes is sufficient to induce apparent viscosity and internal friction, even in the harmonic limit of interatomic interactions, i.e. without the need of anharmonic effects such as local instabilities, plasticity, or other kinds of irreversible structural changes (Fig.5-right). The interatomic bonds here remain unchanged. This kind of harmonic dissipative process is dominant at very high frequencies, and it is not very sensitive to the temperature.

\begin{figure}[tbp]
\includegraphics[width=0.3\linewidth]{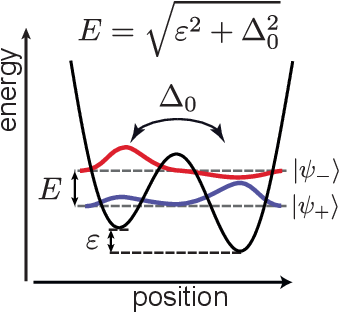}
\includegraphics[width=0.3\linewidth]{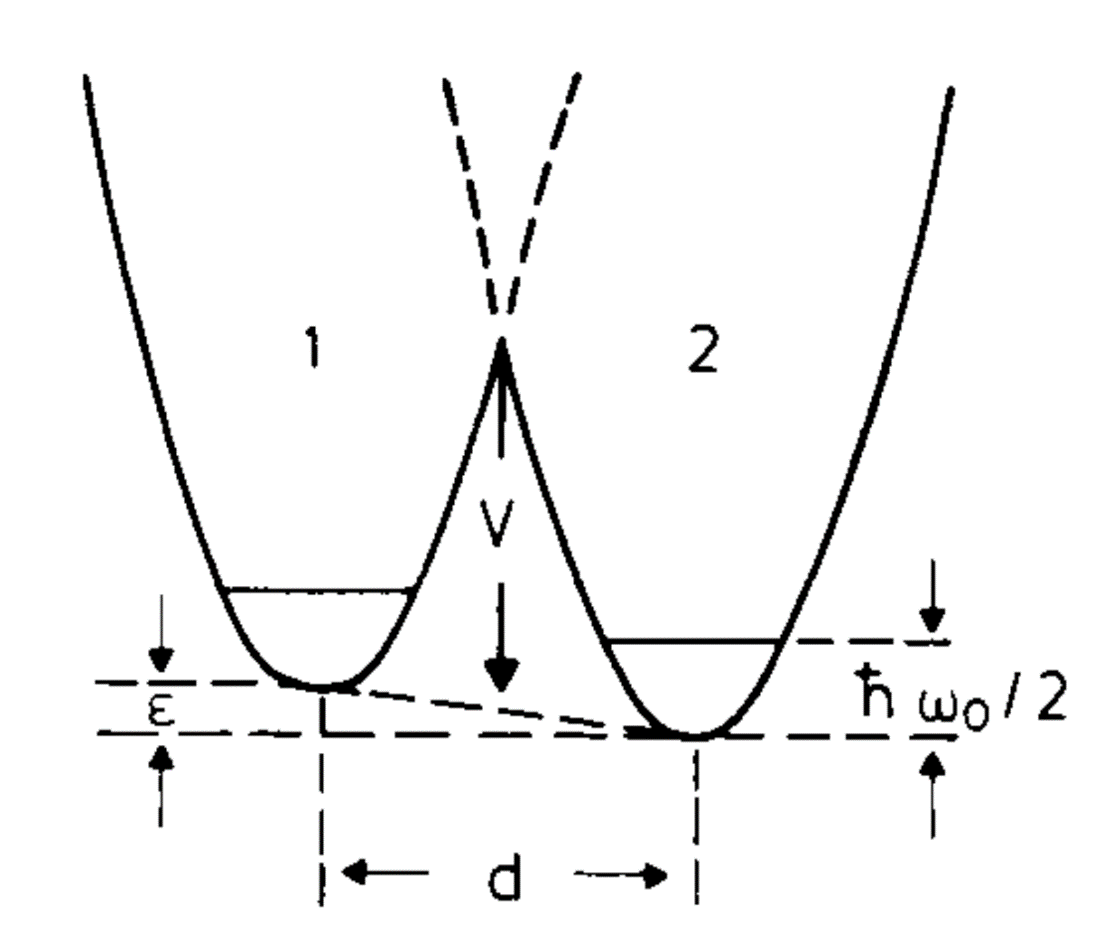}
\includegraphics[width=0.37\linewidth]{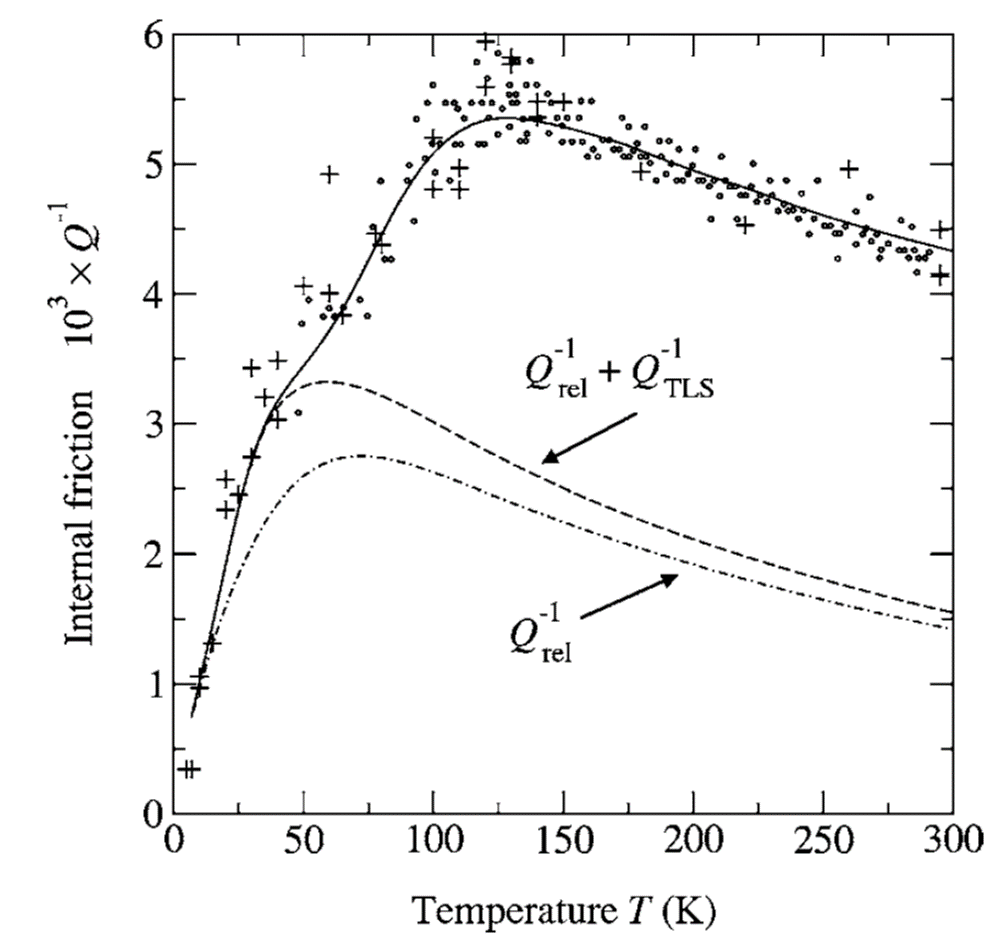}
 \caption{Illustrations of thermal sensitive dissipation mechanisms in the low temperature-low frequency regime: (a) two-level systems (TLS) with tunneling~\cite{Muller2019} (b) double-well potential with thermal activation and thermally activated relaxation (TAR)~\cite{HunklingerReview} (c) contribution of TAR and of TLS to the dissipation measured with Brillouin spectroscopy in silica glass~\cite{Vacher2005}.}
 \label{fig:LowT}
 \end{figure}

\subsection{Anharmonic processes for acoustic attenuation}
Other kinds of dissipative effects have however been reported in the litterature, that are sensitive to the temperature. It is the case of double well potential effects, and two-level systems that seems to play an important role in the low-frequency / low temperature cases. They already have been reviewed \CORR{in details} in~\cite{HunklingerReview}\CORR{, and there are now new attemps to identify them in real systems at the microscopic level from numerical simulations~\cite{Damart2018, Fan2014, Ciarella2023, Mocanu2023} as well as from experimental devices~\cite{Muller2019}}. \CORR{They are related to a transition from one stable configuration to another one, separated by some energy barrier. The general observation is that, for $MHz$ excitations (ultrasonic regime), the inverse mean-free path, or acoustic attenuation, shows two peaks as a function of the temperature: one is located around $50K$ and the other around $5K$. Two categories of related processes have been identified depending on how the system overcomes the energy barrier.}

The mean-free path resulting from a relaxation process is expressed as~\cite{Jackle1976}
\begin{equation}
\ell^{-1}_{MFP}=\frac{A}{T}\int_0^\infty f(V)\frac{\omega^2\tau(V)}{1+\omega^2\tau^2(V)},dV
\label{ell-TAR}
\end{equation}
with $\omega$ the frequency, $T$ the temperature, and $f(V)$ the distribution of the activation energies $V$. \CORR{This expression results simply from the rate equation in the occupation number of the two equilibrium states, taking into account the frequency dependent elastic field in the free energy of the system. The mean-free path then results from the imaginary part of the resulting effective elastic modulus as detailed in~\cite{Jackle1976} and remined in Eq.(~\ref{eq:IntFri}).} For $\omega\tau\ll 1$ the attenuation is proportional to $\omega^2$, and it is constant $\propto 1/\tau$ in the opposite limit. Such a relaxation process is acting in double well potentials (see figure 6-a), and is related to the crossing of the energy barrier $V$ with thermal activation \CORR{and acoustic excitation}, and the subsequent thermal relaxation \CORR{at a time scale equal to $\tau$} in the new equilibrium position. It is also refered to as "Thermally activated Relaxation" (TAR).

In the very low temperature regime, another process in involved, because in this temperature range the transition from one equilibrium state to another may be due to quantum tunneling effect. In this case (figure 6-b), the quantities that matter are \CORR{the energy splitting between the two levels $E=\sqrt{\epsilon^2+\Delta^2}$ with $\Delta/\hbar$ the tunneling frequency depending on the barrier height, the distance, and the mass of the structure involved in the transition,} compared to the incident energy $\hbar\omega$ of the vibration. This process is called "Resonant absorption" or "Two-level system" (TLS). \CORR{The difference with the previous case (Eq.~\ref{ell-TAR}), is that now the relaxation time $\tau$ is given by the quantum Golden rule~\cite{Jackle1976}, including the coupling to  the elastic wave in the context of phonon-assisted tunneling.} In this case, the resulting mean-free path is written~\cite{Phillips1987,HunklingerReview}
\begin{equation}
\ell^{-1}_{MFP}=B\omega\frac{\tanh (\hbar\omega/2k_BT)}{\left(1+J/J_c\right)^{1/2}}
\label{ell-TLS}
\end{equation}
with $J$ the acoustic intensity, yielding to $\ell^{-1}\propto\omega$ for thermal phonons (when $\hbar\omega\gg 2k_BT$).

These two phenomena (TAR and TLS) are not sufficient to explain the large dissipation observed thanks to Brillouin scattering in the $GHz$ to $THz$ range (hypersonic regime) as seen in the figure 6-c. \CORR{The fit, shown in this figure, has been obtained by combining relative velocity change measurements to get the parameters charaterizing the distribution of double-well potentials, and Brillouin broadening~\cite{Vacher2005}. To explain the discrepancy between Brillouin measurements and theoretical predictions in this frequency range, R. Vacher et al. refered to "relaxation due to anharmonic interactions between thermal phonons perturbed by the acoustic excitation"~\cite{Vacher2005}. It is however not clear wether this mechanism is not mainly due to the one detailed in the previous part in the harmonic regime. Finally,} additional anharmonic effects (with crossing of larger energy barriers) involving a large set of competing equilibrium positions could also contribute and show a temperature dependence of the acoustic attenuation at higher temperatures. \CORR{Note however, that no anharmonic effect has been observed in the very high frequency range (above the Ioffe-Regel frequency) as discussed in~\cite{Damart2017}.}

\section{Temperature dependence of the Thermal conductivity}

In amorphous materials, the thermal conductivity is very low compared to crystals (see for example the case of silicon in Figure 2-Bottom). Moreover, it seems, that the temperature sensitivity of the thermal conductivity behaviour of every amorphous materials share common features like a sub-$T^3$ dependence in the low temperature limit, followed by a plateau near $10 K$ and then a monotonous increase up to the melting temperature. To understand this behaviour, we review in this part different microscopic interpretation of the thermal conductivity. In general heat carriers are electrons and phonons. The eletronic contribution to the thermal conductivity refers usually to the Wiedemann-Franz law~\cite{Kittel2004}  
\begin{equation}
\kappa=L_0 T\sigma
\end{equation}
with $\sigma$ the electrical conductivity and $L_0$ the Lorenz constant. In insulators (vs. semi-conductors) like dielectric glasses, this contribution is negligible (vs. comparable) with respect to the phononic contribution. We focus in the next parts only on the phononic contribution to the thermal conductivity.

\subsection{Kinetic Theory of Thermal Conductivity}

The most used theory for the thermal conductivity is the kinetic theory~\cite{Kittel2004}. This theory is based on the existence of a mean-free path $\ell_{MFP}$ for the heat carriers. The Fourier law postulates the existence of a linear relation between the temperature gradient and the heat flux (heat transfer per unit surface and unit time). For example in the $x$-direction:
\begin{equation} 
j_q=\frac{\delta Q_x}{dSdt}=-\kappa_x\frac{\partial T}{\partial x}
\end{equation}
Considering that the heat variation is directly related to the internal energy $\delta Q_x=dU=C_v dT$, the heat flux can be rewritten 
\begin{equation}
j_q=-\frac{N}{\Omega}\langle v_x C_v dT \rangle
\end{equation} 
with $N/\Omega$ the number of heat carriers per unit volume, $\Omega$ is the volume of the sample. Replacing 
\begin{equation}
dT = \frac{dT}{dx}\ell_{MFP}
\end{equation}
and assuming the isotropy in the velocity of phonons, the expression for the heat flux becomes 
\begin{equation}
j_q\approx-\frac{1}{3}\frac{N}{\Omega}C_v\vert\vert v_g\vert\vert\ell_{MFP}\frac{dT}{dx}
\end{equation} 
that gives the usual description $$\kappa=\frac{1}{3}\frac{N}{\Omega}C_v\vert\vert v_g\vert\vert\ell_{MFP}$$
However, each frequency $\omega$ contributes to the heat conductivity. The general formula, restricted to the propagative phonons characterized by a well defined mean-free path $\ell_{MFP}$, that is below the Ioffe-Regel frequency $\omega_{IR}$ is thus
\begin{equation} 
\kappa^{prop} = \frac{1}{3}\int_0^{\omega_{IR}} C_v(T,\omega) v_g(T,\omega)\ell_{MFP}(T,\omega)g(\omega)d\omega \label{eq:kappaProp}
\end{equation}
where $g(\omega)$ is the density of vibrational states. Above the Ioffe-Regel frequency, the mean-free path is ill-defined, but the transportation of energy is ensured by the diffusivity. The diffusivity $D$ appears as well in the Heat equation
\begin{equation}
\frac{\partial T}{\partial t}=D\Delta T + \frac{\Omega}{C_v} q_{source}
\end{equation}
where $q_{source}$ are heat sources per unit volume. Diffusivity can be defined as $D=\kappa \Omega/C_v$ but mainly appears as the diffusion coefficient for the spreading of temperature, that is of the atomic kinetic energy - or velocity fluctuations - inside the system. Indeed, the kinetic definition of the temperature is \begin{equation}
\left\langle \frac{1}{2}m\vert\vert v-\langle v\rangle\vert\vert^2 \right\rangle = \frac{3}{2}k_BT
\end{equation}
The diffusivity in the heat equation is thus strictly equivalent to the diffusivity infered from the spreading of the kinetic energy of the wave-packets shown in Fig.1 third-right. Considering the diffusive spreading of thermal changes in the system, the equation defining the heat conductivity must obviously be changed in
\begin{equation}
\kappa^{diff} = \int_{\omega_{IR}}^{\omega_{max}} C_v(T,\omega) D(\omega) g(\omega)d\omega
\label{eq:kappaDiff}
\end{equation}
Note that this expression has already been used in atomistic simulations~\cite{Larkin2014}, but is unfortunately rarely mentioned in the litterature, given the weight - which seems to us exaggerated - attributed to the mean free path in the interpretation of the energy dissipation.

\subsection{Dedicated theories for thermal conductivity in amorphous materials}

Vibrations in solids result from an extended collective motion, and the Phonon Gas Model is generally a poor approximation of the dynamics of the underlying waves. In the Phonon Gas Model~\cite{Kittel2004}, the density of phonons is low and phonons - which are defined by their wavevector- are supposed to interact punctually through short time collisions~\cite{Kreuzer1981}. In amorphous materials, an additional obstacle comes from the difficulty already discussed of associating a vibrational excitation with a single wave vector\cite{Beltukov2016}. 
In the description of energy transport based on the dynamics of phonons with well-defined wave-vector, the interactions between phonons are thus essential. There are two methods used to take account of these interactions: the Bogolioubov-Born-Green-Kirkwood-Yvon (BBGKY) hierarchy to go beyond the Boltzmann Transport Equation~\cite{Kreuzer1981}, and the Allen and Feldman method based on the Kubo approximation for the heat conductivity that takes into account the non-diagonal terms of the heat current operator~\cite{Allen1993}. 

The BBGKY hierarchy is based on the general equations for the $n$-particles distribution function and their use in energy conservation equations. The $n$-particles distribution function $f(r_1,p_1,…,r_N,p_N,t)$ satisfies the Liouville equations~\cite{Kreuzer1981} 
\begin{equation}
\frac{\partial f}{\partial t}+\sum_{i=1}^N \left(\frac{\partial f}{\partial r_i}.\frac{dr_i}{dt}+\frac{\partial f}{\partial p_i}.\frac{dp_i}{dt}\right)=0=\frac{\partial f}{\partial t}+\{f,H\}
\end{equation}
with the Hamiltonian $H$ including the kinetic and the potential energy of the interacting particles 
\begin{equation}
H=\sum_{i=1}^N \frac{p_i^2}{2m}+U_{pot}(r_i )+\sum_{i<j} u(r_i-r_j )
\end{equation}
For the one-particle distribution $f_1 (r_1,p_1,t)$, the Boltzmann equation is replaced by the exact relation:
\begin{equation}
\left[\frac{\partial}{\partial t}+\frac{p_1}{m}\frac{\partial}{\partial r_1}+F(r_1)\frac{\partial}{\partial p_1}\right] f_1 (r_1,p_1,t)=\int dr_2 dp_2  \frac{\partial u(r_1-r_2)}{\partial r_1}\frac{\partial}{\partial p_1} f_2 (r_1,p_1,r_2,p_2,t)
\end{equation} 
where the right-hand side replaces the collision term of the Boltzmann equation. This yields to the energy conservation equation
\begin{equation}
\frac{\partial u^{tot}}{\partial t}+\nabla .\left(u^{tot} {\bf v}+{\bf j}_q \right)=-\left(P^{cin}+P^{dyn} \right):\left(\nabla {\bf v} \right)
\end{equation}
with the kinetic and dynamic pressures $P^{cin}$ and $P^{dyn}$, the first one being related to the velocity fluctuations and the last one being related to the interparticle interactions including the 2-particles distribution function (see Ref.~\cite{Kreuzer1981} for more details), and with the heat flux
\begin{equation}
j_q=j_q^{cin}+j_q^{pot}
\end{equation}
combining two contributions to the heat flux: the usual $j_q^{cin}$ corresponding to the kinetic single-phonon contribution to the heat flux, or thermal motion of the phonons, together with the $j_q^{pot}$ term corresponding to heat conduction via the two-body interaction of the molecules. After a linear expansion of the one-particle distribution function around its equilibrium value, this last term yields a contribution to the thermal conductivity that is related to the stiffness of the interatomic interactions~\cite{Kreuzer1981}, highlighting the important role of thermo-mechanical couplings in condensed matter.  

On another side, the Allen and Feldman theory of Thermal Conductivity~\cite{Allen1993} shows the crucial role of “diffusons” on the thermal conductivity in amorphous solids. Their calculation is based on the Green-Kubo formula for the thermal conductivity as a function of the Heat flux operator, written in the normal modes basis. 
In this formalism, the heat current is obtained as $tr(\rho S)$, with $\rho$ the \CORR{density matrix} and $S$ the Heat current operator. \CORR{In classical mechanics, the main contribution to the heat current operator in solids~\cite{Allen1993,Hardy1963} is due to atomic oscillations around equilibrium positions. It can be rewritten as}
\begin{equation}
{\bf S} = \frac{1}{2\Omega} \sum_{l,n}\sum_{\alpha,\gamma}\left({\bf R}_l-{\bf R}_n \right) \frac{p_l^\alpha}{m_l} \frac{\partial^2 E}{\partial u_l^\alpha\partial u_n^\gamma} u_n^\gamma
\end{equation} 
that is a function of the dynamical matrix components, with $l$ and $n$ standing for the atoms, $u$ the displacements, \CORR{ $\Omega$ the volume,} $\alpha$ and $\gamma$ the spatial directions. The Green-Kubo relation results from the departure from equilibrium of the local temperature T(x). \CORR{The average temperature being noted $T$ with $\beta=1/k_BT$}, it gives  the following expression for the heat conductivity
\begin{equation}
\kappa_{\mu\nu} = \frac{\Omega}{T} \int_0^\beta d\lambda \int_{-\infty}^0 dt \langle e^{\lambda H} S_\mu (t) e^{-\lambda H} S_\nu (0)\rangle
\end{equation} 
or, with the quantum notations allowing an easier calculation~\cite{Allen1993}:
\begin{equation}
\kappa_{\mu\nu}(\omega) = \frac{\Omega}{T} \int_0^\beta d\lambda \int_{-\infty}^0 dt e^{i(\omega+i\eta)t} \sum_{m,n}\frac{e^{-\beta E_n }}{Z} \langle n\vert e^{\lambda H} S_\mu (t) e^{-\lambda H} \vert m\rangle\langle m\vert e^{iHt/\hbar} S_\nu (t) e^{-iHt/\hbar} \vert n\rangle
\end{equation}
Considering for example the diagonal term $\kappa=Re\left(\kappa_{xx}\right)$ and setting $S_x=S=\sum_{i,j}S_{ij} a_i^+ a_j$, it gives two contributions  \CORR{within the harmonic approximation}: 
\begin{equation}
\kappa^I (\omega) = \frac{\pi}{\Omega T}\sum_{i\ne j}\left[ \frac{\langle n_j \rangle-\langle n_i \rangle}{\hbar (\omega_i-\omega_j)}\right] \vert S_{ij} \vert^2 \delta(\omega_i-\omega_j-\omega)
\end{equation}
and
\CORR{\begin{equation}
\kappa^{II} (\omega) = \frac{\beta}{\Omega T} Re\left[\frac{i}{\omega+i\eta}\right] \sum_{i,j}\langle n_i n_j \rangle S_ {ii} S_{jj} \approx \frac{1}{\hbar T\Omega}Re\left[\frac{i}{\omega+i\eta}\right] \sum_i \left[-\frac{\partial \langle n_i \rangle}{\partial \omega_i} \right] S_{ii}^2
\end{equation}}
The second contribution $\kappa^{II}$ is the “intraband” conductivity. It is equivalent to the Drude model for electrons in metals, and gives at $\omega=0$ the well known relation \CORR{$\kappa_{\mu\nu}^{II}=\sum_i v_i^\mu v_i^\nu \tau_i C_i$ or $\kappa_{\mu\nu}^{II}=1/3 Cv\ell\delta_{\mu\nu}$} when a cubic symmetry holds, with $C_i=\frac{\hbar \omega_i^2}{\Omega T}\left[-\frac{\partial \langle n_i \rangle}{\partial \omega_i}\right]$ being the specific heat, \CORR{and $\tau_i=1/\eta$ ($\eta\rightarrow 0$) being an empirical broadening parameter due to collisions or finite mean-free path}. $\kappa^{II}$ is the only term contributing to the thermal conductivity in pure crystals in the quasi-harmonic approximation, and it is zero when the phonons are not ballistic ($S_{ii}=0$). On the contrary, the first contribution $\kappa^I$ is the “interband” conductivity. It is related to out-of-diagonal terms in the Heat current operator, that are transport terms involving the coupling between different phonons (wave packets) associated to different wave vectors. In the low frequency limit ($\omega\rightarrow 0$), it is written 
\begin{equation}
\kappa^I=\frac{1}{\Omega} \sum_i C_i (T)D_i
\label{eq:Allen}
\end{equation}
with $D_i$ the diffusivity of the mode $i$, $D_i=\frac{\pi \Omega^2}{3\hbar^2 \omega_i^2} \sum_{j\ne i} \vert S_{ij}\vert^2 \delta(\omega_i-\omega_j )$. This expression is fully in agreement with the kinetic theory in the diffusive regime. A non-zero value of $S_{ij}$ implies that the vibrations are extended and not localized. This expression confirms that {\it{localized}} non-propagating vibrations will not contribute to the thermal conductivity. The contribution of the diffusivity $D_i$ to the thermal conductivity is in general absent in crystals, but it appears here as the contribution of non ballistic but extended vibrations to the thermal conductivity. As shown before, these vibrations, excited in the frequency range of “diffusons”, are present on a large set of frequencies in amorphous materials. It is very important to underline here the non-local character of diffusons, that are defined as extended and harmonic vibrations, but without plane wave character~\cite{Allen1993} due to disorder. This non-local character is the condition to contribute to the thermal conductivity. \CORR{These methods have been implented in recent numerical calculations of the thermal conductivity of amorphous materials compared to crystals~\cite{Isaeva2019,Simoncelli2023}}

\subsection{Harmonic Wave-Packets contribution to the thermal conductivity in glasses}

In this part, we will discuss the temperature dependence of the thermal conductivity given by the kinetic theory in the propagative, and in the diffusive regimes. In general, the contribution of the different eigenfrequencies to the heat capacity is given by the Statistical \CORR{Physics} derivation from the partition function
\begin{equation}
C_v(\omega,T) = \frac{N}{k_B\Omega}\left(\frac{\hbar\omega}{k_B T}\right)^2\frac{\exp\left(\frac{\hbar\omega}{k_B T}\right)}{\left(\exp\left(\frac{\hbar\omega}{k_B T}\right)-1\right)^2}
\label{eq:Cv}
\end{equation}
In the Debye model with the approximation $g(\omega)\propto\omega^2$ used up to the Debye frequency $\omega_D$, the global heat capacity $C_v$ resulting from the integration over all the frequencies $\omega$ of $C_v(\omega)g(\omega)$ gives $C_v\propto T^3$ when $k_BT\ll\hbar\omega_D$ and $C_v\propto 3Nk_B$ when $k_BT\gg \hbar\omega_D$.

We have seen in the previous parts, that there are two characteristics of amorphous materials compared to crystals: first the existence of “diffusons” that are normal modes giving rise to diffusive propagation of plane waves at the same frequency, and secondly the unusual frequency dependence of the mean-free path of phonons.
Specifically, the “anomalous” temperature dependence of the heat conductivity in amorphous materials is explained in the low frequency regime by the frequency dependence of the mean-free path. There are different explanations and different frequency dependences for the mean-free path of phonons. The most famous one is the “Tunneling Two-level Systems” (TLS)~\cite{Phillips1987} already discussed before. In this case, we have seen that, in the low acoustic intensity regime, the mean-free-path scales like $ \propto 1/\omega\propto 1/T$ as a result of the Fermi Golden-rule for the attenuation related to the transition between energy states separated by an energy $\hbar\omega$. This $\omega$-dependence of $\ell_{MFP}$  thus lowers the exponent in the temperature dependence of the thermal conductivity compared to the constant-$\ell_{MFP}$ case, as a result of the kinetic expression of the heat conductivity in the propagative (ballistic) regime. Specific temperature dependence of the heat capacity may also play a role~\cite{Churkin2021}. Another well-known dependence of the mean-free path results from the Rayleigh scattering on structural disorder in the low-frequency regime~\cite{Gelin2016}. In this case, the mean-free path scales as $\propto 1/\omega^4$ with possible logarithmic corrections if the anisotropic quenched stresses resulting from the quenching process matter~\cite{Gelin2016}. The resonant interaction between vibrational soft modes and sound waves~\cite{Buchenau1992} has been calculated as giving rise to another frequency dependence of the mean-free path scaling like $\propto 1/\omega^3$. At finite temperature, Molecular Dynamics simulations~\cite{Mizuno2020,Larkin2014} have also measured an inverse acoustic attenuation (proportional to the mean-free path) scaling like $\propto 1/\omega^{(3/2)}$~\cite{Mizuno2020} and also $\propto 1/\omega^2$~\cite{Larkin2014} in the sonic to ultrasonic frequency range. This scaling is sensitive to the temperature. 

We have performed classical Molecular Dynamics simulations of amorphous silicon described with harmonic interactions at very low temperature~\cite{Beltukov2016,Tlili2019,Desmarchelier2021}. In these simulations, we have computed the thermal conductivity from the kinetic theory (equation~(\ref{eq:kappaProp}) in the ballistic regime) using the expression~(\ref{eq:Cv}) for the frequency dependent Heat capacity $C_v(\omega,T)$. We have also computed the mean-free path in the very low temperature regime by considering either the attenuation of the enveloppe of a wave-packets excitation, either the \CORR{broadening} of the dynamical structure factor that is shown to give similar results. We have not considered the effect of temperature on the mean-free path, since we have chosen to focus only on the harmonic behaviour. In our case, the measured mean-free path is thus independent on the temperature, but frequency dependent. We have shown interestingly that the resulting low temperature dependence of the thermal conductivity follows an approximate power-law behaviour that could be fitted as $\propto T^2$ below $40 K$, in agreement with the experiments~\cite{HunklingerReview}. In this model calculation, this anomalous behaviour results only from the non-trivial $\omega$-dependence of the mean-free path obtained either from the dynamical structure factor, or from the apparent attenuation of the enveloppe of the kinetic energy of the wave-packets at sufficient low frequencies~\cite{Beltukov2018}. 

In the high frequency regime (above the Ioffe-Regel frequency), a scaling of the mean-free path $\propto 1/\omega^2$ is reported in the experimental litterature~\cite{Baldi2010,Damart2018,Luo2022}, but in this regime the Damped Harmonic Oscillator fits used to identify the mean-free path and the acoustic attenuation are no \CORR{longer} valid~\cite{Baldi2010,Damart2017,Beltukov2018}. As shown before, the mean-free path indeed becomes ill-defined in this frequency range. The ballistic contribution to the heat conductivity thus saturates above the Ioffe-regel frequency, that is for temperatures larger than $\hbar\omega_{IR}/k_B\approx 50 K$. This saturation could be at the origin of the plateau in the temperature dependence of the thermal conductivity.

Finally, the diffusivity and the Allen and Feldman model for the thermal conductivity - Eq.~(\ref{eq:kappaDiff}) - allows explaining the monotonous increase of the thermal conductivity in the high temperature regime. This is due to the cumulated contribution of diffusons to the thermal conductivity up to the melting temperature regime in glasses~\cite{Tlili2019}. In amorphous silicon, interestingly, the global contribution of diffusons dominates the calculation of heat conductivity in amorphous silicon~\cite{Tlili2019}. The diffusive effect seems enhanced by the temperature increase simply due to the cumulative contribution of the higher frequencies, that is not opposed by the thermal decay of the mean-free path, since this one is ill-defined here.

It is interesting to compare now pure amorphous materials and glass-ceramics or glasses nanostructured with crystalline inclusions (see figures 4, 7 and 8 for a comparison between different kinds of nanostructurations). In the case of spherical crystalline inclusions, the presence of interfaces diminishes the Ioffe-Regel frequency and thus the ballistic contribution to the thermal conductivity, especially when the acoustic mismatch between the matrix and the inclusions is higher. But for the same reasons, the diffusive contribution to the thermal conductivity starts at smaller $\omega_{IR}$ frequencies, and becomes consequently more important than in pure crystals. The additional contribution of the ballisc and diffusive parts gives rise to equivalent values for the thermal conductivity, with a small increase of the thermal conductivity for stiffer inclusions due to a its ballistic contribution (higher sound velocity). The ballistic contribution is in general negligible compared to the diffusive one in all these cases.

A more significant variation of the thermal conductivity can however be obtained in two limiting cases. The first one is when the \CORR{nanostructure} is made of voids. It is the only case where the diffusive contribution to the thermal conductivity is decayed compared to the pure amorphous sample. The resulting global conductivity of the porous medium is smaller than the amorphous one (see figure 8-c). The second case results from the percolation of inclusions in the system. In this opposite case, the ballistic contribution to the thermal conductivity is strongly enhanced and even clearly dominates the diffusive one~\cite{Desmarchelier2021}. In this last case, the thermal conductivity is controlled by the ballistic transport of heat taking place inside the crystalline percolating path. Playing with the geometrical arrangement of the amorphous phase on the crystalline wiskers, anisotropic and even single-way heat transport may be infered in the nanodesigned materials~\cite{Desmarchelier2021}. 

\begin{figure}[tbp]
\includegraphics[width=0.65\linewidth]{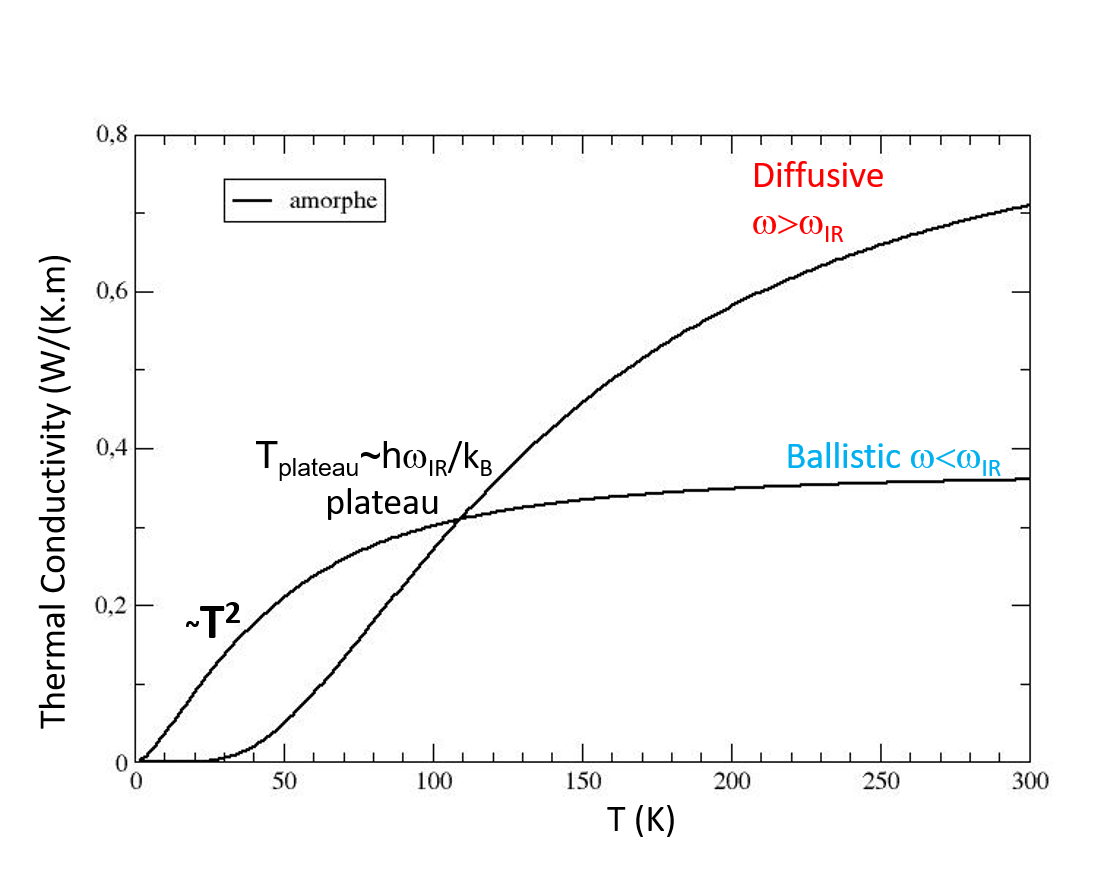}
\includegraphics[width=0.65\linewidth]{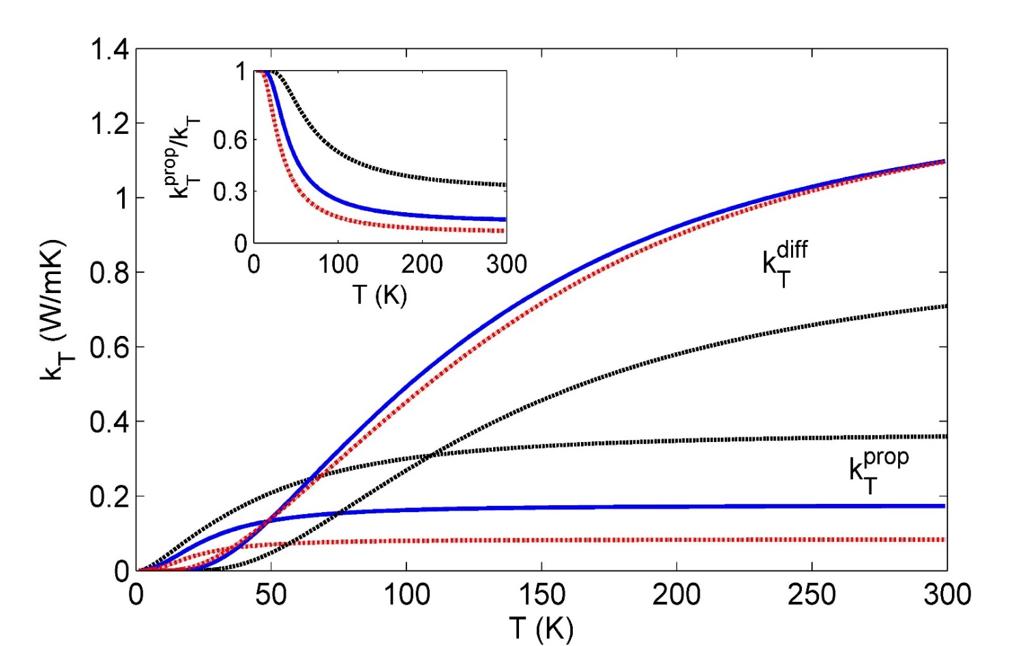}
\caption{Ballistic and Diffusive contributions to the thermal conductivity in: (a) amorphous Si sample (b) amorphous Si sample with spherical crystalline inclusions with radius $25\AA$ (from Ref.~\cite{Tlili2019})}
\end{figure}

\begin{figure}
\includegraphics[width=1\linewidth]{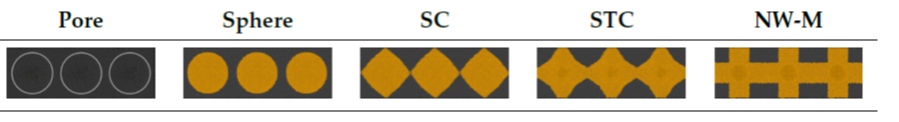}
\includegraphics[width=1\linewidth]{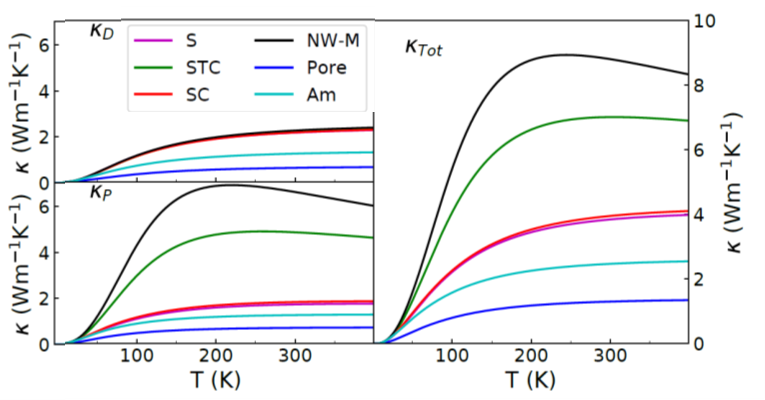}
\caption{ top: Different sets of crystalline nanoinclusions: Pores, Amorphous (Am), Spherical (S), Conical touching at the edge (SC), Conical with overlap (STC), nanowires with amorhpous matrix (NW-M) bottom: Ballistic and Diffusive contributions to the thermal conductivity for the nanostuctured samples. See~\cite{Desmarchelier2021} for more details.}
 \label{fig:ThCond}
 \end{figure}

\section{Conclusion}

To conclude, we have shown in this paper that there are different possible interpretations for the temperature dependence of the Thermal conductivity in glasses. We focused here on the phononic contribution. The difficulty in converging towards a common interpretation at the microscopic scale, is linked to the difficulty of agreeing on the notion of "phonon" in a medium where the acoustic waves are not uniquely defined by a wave vector, and by the variety of dissipation mechanisms in a varied energy landscape with a wide distribution of energy barriers. We have proposed here a short  review of these processes. One source of confusion is related to the fact that harmonic vibrations in glasses cannot be described by a single wave-vector. The discrepancy with single wave-vectors excitations yields to apparent dissipation of phonons, mainly related to local redirection of acoustic excitations in the harmonic approximation. The main message for the harmonic processes, is that the description of the apparent acoustic attenuation and heat transfer with a mean-free path is not valid in the hypersonic range (THz) of diffusons. In this frequency range, the concept of mean-free path must be replaced by the notion of diffusivity, that cannot be related seriously to a mean-free path since the velocity of sound and of acoustic excitations (wave-packets) is ill-defined in this frequency range. This specific behaviour results from strong scattering of acoustic waves on the mechanical heterogeneities of amorphous materials at the nanometer scale. The two harmonic contributions, in the low scattering - ballistic or propagative regime (for $\omega < \omega_{IR}$) and in the strong scattering regime (for $\omega>\omega_{IR}$) contribute separately to the thermal conductivity. The later is sufficient to explain the anomalous temperature dependence in the low temperature regime (due to the disorder-induced frequency dependence of the mean-free path) and the plateau of the thermal conductivity when its contribution saturates, while the last one explains the monotonous increase of the thermal conductivity up to the melting temperature. The combination of both contributions is sufficient to understand the role of structural design at the nano-scale, especially why a percolation of crystalline inclusions allows increasing the thermal conductivity by boosting its ballistic contribution, and why the use of voids allows decreasing it by decreasing strongly the acoustic diffusivity. 

Another contribution to the thermal conductivity is due to anharmonic effects like thermal relaxation phenomena in two-level systems, or equivalently hysteretic behaviour due to crossing of energy barriers. This contribution yields a new dependence of the thermal conductivity on the frequency and on the temperature. This contribution may dominate the low frequency regime when the temperature is sufficiently high~\cite{Mizuno2020}, but it becomes also sensitive to the dissipation processes~\cite{Damart2017}. Anharmonic effects are visible mainly in the low frequency regime, because they are related to low energy barriers. Note that in the solid state, the temperature activates irreversible dissipative rarrangements whose density may affect the frequency dependence of the acoustic attenuation and thus of the mean-free path. But below the glass transition temperature, thanks to the elastic couplings, it is not able to decrease sufficiently the mean-free path to prevent the propagation of acoustic waves.

Finally, the nanostructuration of the glasses with crystalline inclusions can help controlling the heat flux by enhancing its ballistic contribution, while a nanostructuration with pores (voids) helps decreasing its diffusive contribution. In the ballistic regime, the mean-free path of phonons is a non monotonous function of the size of the nanocristalline inclusions, due to complex vibrations taking place in nano-composites, like resonant vibrations of inclusions or gallery waves. Anderson's localization due to disorder would be the only way to strictly stop heat transfer, but it is restricted to few and tiny frequency ranges at which the glass is also an acoustic insulator, in such a way that heat transfer will be supported by neighbouring frequencies. 

The wave nature of heat transfer~\cite{Moon2019}, but also the difference in the behaviour of wave packets in the crystals and in the amorphous materials for nanometer wavelengths helps however managing heat in materials. For example, it is possible to induce rectification (one-way heat transfer) by playing with the thickness of amorphous layers~\cite{DesmarchelierRec21}. It is also possible to imprint Young slits and thus to produce wave interferences and local heating in a crystal using amorphous inclusions~\cite{Desmarchelier2023}. Possibilities offered by nanoscale design of thermal circuits~\cite{TherMan} are vast.

\section*{Acknowledgements}
This work was encouraged and strongly supported by the doctoral theses of Paul Desmarchelier and Haoming Luo, under shared supervisions with Valentina Giordano, Anthony Gravouil and Konstantinos Termentzidis. Paul Desmarchelier performed the atomistic calculations of thermal conductivities. I thank Yaroslav Beltukov for his efficient contribution to our common work, and S\'ebastien Auma\^{i}tre for his constant support.

\bibliographystyle{crunsrt}
\bibliography{Article-Verres-Tanguy}

\end{document}